\documentclass[aps,prd,preprint,,groupptaddress,tightenlines,nofootinbib]{revtex4}
\long\def\symbolfootnote[#1]#2{\begingroup%
\def\thefootnote{\fnsymbol{footnote}}\footnote[#1]{#2}\endgroup}

\usepackage{amsfonts,amsmath,amsthm,amssymb}
\usepackage{mathrsfs}
\usepackage{bm}
\usepackage{color}
\usepackage{axodraw}
\usepackage{epsfig}

\newcommand{\PRE}[1]{{#1}}   
\newcommand{\met} {\not\!\! E_T}

\newcommand{\postscript}[2]{\setlength{\epsfxsize}{#2\hsize}
   \centerline{\epsfbox{#1}}}
\newcommand{\comment}[1]{}

\newcommand{\el}[1]{\label{#1}}
\newcommand{\er}[1]{\eqref{#1}}

\newcommand{\ci}[1]{}

\newcommand{\ke}{\rangle}
\newcommand{\br}{\langle}

\newcommand{\p}{\partial}

\newcommand{\ba}{\begin{eqnarray}}
\newcommand{\ea}{\end{eqnarray}}
\newcommand{\be}{\begin{equation}}
\newcommand{\ee}{\end{equation}}
\newcommand{\bay}[1]{\left(\begin{array}{#1}}
\newcommand{\eay}{\end{array}\right)}

\def\CD{{\cal D}}

\begin{document}

\vspace{2cm}

\title{
\PRE{\vspace*{1in}}
$\bm{U(3)_C \times Sp(1)_L \times U(1)_L \times U(1)_R}$
}

\author{Luis Alfredo Anchordoqui}
\affiliation{Department of Physics, University of Wisconsin-Milwaukee,
P.O. Box 413, Milwaukee, WI 53201, USA
}

\date{August 2011}

\PRE{\vspace*{.5in}}

\begin{abstract}\vskip 3mm
\noindent We outline the basic setting of the $U(3)_C \times Sp(1)_L \times U(1)_L \times U(1)_R$  gauge theory and review the associated phenomenological aspects related to experimental searches for new physics at hadron colliders.

\end{abstract}

\maketitle

\section{General Idea} 

The Standard Model (SM) 
is a spontaneously broken Yang-Mills theory with gauge group $SU(3)_C
\times SU (2)_L\times U(1)_Y \, .$ Matter in the form of quarks and
leptons ({\it i.e.}~$SU(3)_{C}$ triplets and singlets, respectively)
is arranged in three families ($i=1,2,3$) of left-handed fermion
doublets (of $SU(2)_L$) and right-handed fermion singlets. Each family
$i$ contains chiral gauge representations of left-handed quarks $Q_i =
(3,2)_{1/6}$ and leptons $L_i = (1,2)_{-1/2}$ as well as right-handed
up and down quarks, $U_i~=~(3,1)_{2/3}$ and $D_i = (3,1)_{-1/3}$,
respectively, and the right-handed lepton $E_i = (1,1)_{-1}$. The
hypercharge $Y$ is shown as a subscript of the $SU(3)_{C} \times
SU(2)_L$ gauge representation $(A,B)$.  The neutrino is part of the
left-handed lepton representation $L_i$ and does not have a
right-handed counterpart.

The SM Lagrangian exhibits an accidental global symmetry $U(1)_B \times U(1)_e \times U(1)_\mu \times U(1)_\tau$, where $U(1)_B$ is the baryon number symmetry, and $U(1)_\alpha$ ($\alpha={e,\mu,\tau}$) are three lepton flavor symmetries, with total lepton number given by $L = L_e + L_\mu + L_\tau$.  It is an accidental symmetry because we do not impose it. It is a consequence of the gauge symmetries and the low energy particle content.  It is possible (but not necessary), however, that effective interaction operators induced by the high energy content of the underlying theory may violate sectors of the global symmetry.

The electroweak subgroup $SU_{L}(2) \times U_{Y}(1)$ is spontaneously
broken to the electromagnetic $U(1)_{\rm EM}$ by the Higgs doublet $H
= (1,2)_{1/2}$ which receives a vacuum expectation value $v\neq0$ in a
suitable potential. Three of the four components of the complex Higgs are `eaten'
by the $W^\pm$ and $Z$ bosons, which are superpositions of the gauge
bosons $W^a_\mu$ of ${SU}(2)_{L}$ and $B_\mu$ of ${U}(1)_{Y}$,
\begin{equation}
W_\mu^\pm=\frac{1}{\sqrt{2}}W_\mu^1\mp \frac{i}{\sqrt{2}}W_\mu^2 
\end{equation}
and
\begin{equation}
Z_\mu=\cos\theta_{W}\,W^3_\mu-\sin\theta_{W}\,B_\mu\,,
\end{equation}
with masses  $ M^2_W = \pi\alpha v^2/\sin^2\theta_{W}$, $M^2_Z = M^2_W/\cos^2\theta_{W},$ and $\alpha\simeq1/128$ at $Q^2~=~M_W^2$. The fourth vector field, 
\begin{equation}
A_\mu = \sin\theta_{W}\,W^3_\mu + \cos\theta_{W}\,B_\mu \,, 
\end{equation}
persists massless and the
 remaining Higgs component is left as a $U(1)_{\rm EM}$ neutral real scalar.  The measured values $M_W\simeq80.4$~GeV and $M_Z\simeq91.2$~GeV fix the weak mixing angle at $\sin^2\theta_{W}\simeq0.23$ and the Higgs vacuum expectation value at $\langle H \rangle = v\simeq246$~GeV~\cite{Nakamura:2010zzi}.

Fermion masses arise from Yukawa interactions, which couple the right-handed fermion singlets to the left-handed fermion doublets and the Higgs field,
\begin{equation}
 {\mathscr L}    =  - Y_d^{ij} \, \bar{Q}_i \, H \, D_j  -  Y^{ij}_u \, \epsilon^{ab} \, \bar Q_{ia} \,  H^\dagger_b \, U_j - Y^{ij}_\ell \, \bar L_i \, H \, E_j +~{\rm h.c.},
\end{equation}
where $\epsilon^{ab}$ is the antisymmetric tensor. In the process of spontaneous symmetry breaking these interactions lead to charged fermion masses,  $m_f^{ij} = Y_f^{ij} \,\, v/\sqrt{2},$ but leave the neutrinos massless~\cite{Halzen:1984mc}.\footnote{One might think that neutrino masses could arise from loop corrections.  This, however, cannot be the case, because the only possible neutrino mass term that can be constructed with the SM fields is the bilinear $\bar L_i L_j^C$ which violates the total lepton symmetry by two units ($L_i^C = C \bar L_i^T$). As mentioned above total lepton number is a global symmetry of the model and therefore $L$-violating terms cannot be induced by loop corrections. Furthermore, the $U(1)_{B-L}$ subgroup is non-anomalous, and therefore $B-L$ violating terms cannot be induced even by nonperturbative corrections. It follows that the SM predicts that neutrinos are \emph{strictly} massless.}  Experimental evidence for neutrino flavor oscillations by the mixing of different mass eigenstates implies that the SM has to be extended~\cite{GonzalezGarcia:2007ib}. The most economic way to get massive neutrinos would be to introduce the right-handed neutrino states (having no gauge interactions, these sterile states would be essentially undetectable) and obtain a Dirac mass term through a Yukawa coupling.

The SM gauge interactions have been tested with unprecedented accuracy,  including some observables beyond even one part in a million~\cite{Nakamura:2010zzi}. Nevertheless, the saga of the SM is still exhilarating because it leaves all questions of consequence unanswered. The most evident of unanswered questions is why there is a huge disparity between the strength of gravity and of the SM forces. This hierarchy problem suggests that new physics could be at play at the TeV-scale, and is arguably {\em the} driving force behind high energy physics for several decades.  Much of the motivation for anticipating the existence of such new physics is based on considerations of \emph{naturalness}. The non-zero vacuum expectation value of the scalar Higgs doublet condensate sets the scale of electroweak interactions. However, due to the quadratic sensitivity of the Higgs mass to quantum corrections from an arbitrarily high mass scale $\Lambda$, with no new physics between the energy scale of electroweak unification and the vicinity of the Planck mass, the bare Higgs mass and quantum corrections have to cancel at a level of one part in $\sim10^{30}$. This \emph{fine-tuned} cancellation seems unnatural, even though it is in principle self-consistent. Thus either the scale of new physics $\Lambda$ is much smaller than the Planck scale or there exists a mechanism which ensures this cancellation, perhaps arising from a new symmetry principle beyond the SM -- minimal supersymmetry (SUSY) is a textbook example~\cite{Wess:1992cp}. In either case, an extension of the SM appears necessary.

In this talk I will discuss the phenomenology of a newfangled extension of the gauge sector, $U(3)_C \times Sp(1)_L \times U(1)_L \times U(1)_R$, which has the attractive property of elevating the two major global symmetries of the SM ($B$ and $L$) to local gauge symmetries~\cite{Anchordoqui:2011eg}.\footnote{The fundamental
  principles of the model are summarized in~\cite{Ibanez:2001nd}. Herein though we replace at full length the  $U(2)_L$ doublets by $Sp(1)_L$ doublets.  Besides the fact that this
  reduces the number of extra $U(1)$'s, one avoids the presence of a
  problematic Peccei-Quinn symmetry~\cite{Peccei:1977hh}, associated in general with the
  $U(1)$ of $U(2)_L$ under which Higgs doublets are
  charged~\cite{Antoniadis:2000ena}.  A point worth noting at this juncture: the compact symplectic group $Sp(1)$ is equivalent to $SU(2)$; our choice of notation will become clear in Sec.~\ref{S5}.} The $U(1)_Y$ boson $Y_\mu$, which gauges the usual electroweak hypercharge symmetry, is a linear combination of the $U(1)$ of $U(3)_C$ gauge boson $C_\mu$, the $U(1)_R$ boson $B_\mu$, and a third additional $U(1)_L$ field $\tilde B_\mu$. The $Q_3$, $Q_{1L}$, $Q_{1R}$ content of the hypercharge operator is given by, \be\el{hyperchargeY} Q_Y = c_1 Q_{1R} + c_3 Q_3 + c_4 Q_{1L} \, ,\ee with $c_1 = 1/2,$ $c_3 = 1/6$, and $c_4 =-1/2$~\cite{Anastasopoulos:2006da}. The corresponding fermion and Higgs doublet quantum numbers are given in Table~\ref{t:spectrum}. The criteria we adopt here to define the Higgs charges is to make the Yukawa couplings ($H \bar U_i Q_i$, $H^\dagger \bar D_i Q_i$, $H^\dagger \bar E_i L_i,$ $H \bar N_i L_i$) invariant under all three $U(1)$'s. From Table~\ref{t:spectrum}, $\bar U_i Q_i$ has the charges ($0, 0, -1)$ and $\bar D_i Q_i$ has ($0, 0, 1$); therefore, the Higgs $H$ has $Q_3 = Q_{1L} = 0$, $Q_{1R} = 1$, $Q_Y = 1/2$, whereas $H^\dagger$ has opposite charges $Q_3 = Q_{1L} = 0$, $Q_{1R} = -1$, $Q_Y = -1/2$. The two extra $U(1)$'s are the baryon and lepton number; they are given by the following combinations: \begin{equation} B=Q_3/3\quad;\quad L=Q_{1L}\quad;\quad Q_Y={1\over 6}Q_3-{1\over 2}Q_{1L}+{1\over 2}Q_{1R}\, ; \end{equation} or equivalently by the inverse relations \begin{equation} Q_3=3B\quad;\quad Q_{1L}=L\quad;\quad Q_{1R}=2Q_Y-(B-L)\, .  \label{bb-l}\end{equation} Even though $B$ is anomalous, with the addition of three fermion singlets $N_i$ the combination $B-L$ is anomaly free.  One can verify by inspection of Table~\ref{t:spectrum}  that these $N_i$ have the quantum numbers of right handed neutrinos, {\em i.e.} singlets under hypercharge. Therefore, this is a first interesting prediction of the $U(3)_C \times Sp(1)_L \times U(1)_L \times U(1)_R$ gauge theory: {\em right-handed neutrinos must exist}.

\begin{table}
\caption{Quantum numbers of chiral fermions and Higgs doublet.}
\begin{tabular}{c|ccccc}
\hline
\hline
 Name &~~Representation~~& ~$Q_3$~& ~$ Q_{1L}$~ & ~$Q_{1R}$~ & ~~$Q_{Y}$~~ \\
\hline
~~$Q_i$~~ & $(3,2)$& $\phantom{-}1$ & $\phantom{-}0 $ & $\phantom{-} 0$ & $\phantom{-}\frac{1}{6}$   \\[1mm]
~~$\bar U_i$~~ & $({\bar 3},1)$ &    $-1$ & $\phantom{-}0$ & $-1$ & $-\frac{2}{3}$ \\[1mm]
~~$\bar D_i$~~ &  $({\bar 3},1)$&    $-1$ & $\phantom{-}0$ & $\phantom{-} 1$ & $\phantom{-}\frac{1}{3}$  \\[1mm]
~~$L_i$~~ & $(1,2)$&    $\phantom{-}0$ &  $\phantom{-}1$ & $\phantom{-}0$ & $-\frac 1 2$  \\[1mm]
~~$\bar E_i$~~ &  $(1,1)$&   $\phantom{-}0$ & $-1$ &  $\phantom{-} 1$ & $\phantom{-} 1$ \\[1mm]
~~$\bar N_i$~~&  $(1,1)$&    $\phantom{-}0$ & $-1$ &  $- 1$ & $\phantom{-} 0$ \\[1mm]
~~$H$~~ & $(1,2)$ &   $\phantom{-}0$     &   $\phantom{-}0$     & $\phantom{-} 1$ & $\phantom{-} \frac{1}{2}$ \\ [1mm]
\hline
\hline
\end{tabular}
\label{t:spectrum}
\end{table}

Before discussing the favorable phenomenological implications of the model, we detail some desirable properties which apply to generic models with multiple $U(1)$ symmetries.\\

\section{Running of the abelian gauge couplings} 

We begin with the covariant derivative for the $U(1)$  fields in the `flavor' $1,\ 2,\ 3, \dots$ basis in which it is assumed that the kinetic energy terms containing $X_\mu^i$ are canonically normalized
\begin{equation}
{\cal D}_\mu = \partial_\mu - i \sum g'_i \, Q_i  \, X_\mu^i \, .
\label{caldmu1}
\end{equation}
The relations between the $U(1)$ couplings $g'_i$ and any non-abelian counterparts  are left open for now. We carry out an orthogonal transformation of the fields
$X_{\mu}^i = \sum_j O_{ij} \, Y_\mu^j$. The covariant derivative becomes
\begin{eqnarray}
{\cal D}_\mu & = & \partial_\mu - i \, \sum_i \sum_j  g'_i \, Q_i  \, O_{ij} \, Y_\mu^j \nonumber \\
 & = & \partial_\mu - i \, \sum_j \, \bar g_j \, \bar Q_j \, Y_\mu^j \,,
\end{eqnarray}
where for each $j$
\begin{equation}
\bar g_j \bar Q_j = \sum_i g'_i \, Q_i \, O_{ij} \, .
\label{gQ1}
\end{equation}
Next, suppose we are provided with normalization for the hypercharge (taken as $j = 1$)
\begin{equation}
Q_Y = \sum_i c_i \,  Q_i \, ;
\label{normalization}
\end{equation}
hereafter we omit the bars for simplicity.
Rewriting (\ref{gQ1}) for the hypercharge 
\begin{equation}
g_Y \,  Q_Y = \sum_i \, g'_i \, Q_i \, O_{i1}
\label{gQ2}
\end{equation}
and substituting (\ref{normalization}) into (\ref{gQ2}) we obtain
\begin{equation}
g_Y \, \sum_i \, Q_i \, c_i = \sum_i \, g'_i \, O_{i1}  \, Q_i .
\label{3}
\end{equation}

One can think about the charges $Q_{i,p}$ as vectors with the components labeled by particles $p$. 
Let us first take the charges to be orthogonal, {\em i.e.} $\sum_p Q_{i,p} Q_{k,p}=0$ for $i\neq k$.  Multiplying (\ref{3}) by $\sum_p Q_{k,p}$,
\begin{equation}
\sum_p Q_{k,p} \, g_Y \, \sum_i \, Q_{i,p} \, c_i = \sum_p Q_{k,p} \, \sum_i \, g'_i \, O_{i1}  \, Q_{i,p} \,,
\end{equation}
we obtain
\begin{equation}
g_Y \, \, c_i =  \, g'_i \, O_{i1}  \,  ,
\end{equation}
or equivalently
\begin{equation}
O_{i1} = \frac{g_Y \, c_i}{g'_i} \, .
\label{tooslow}
\end{equation}
Orthogonality of the rotation matrix, $\sum_i O_{i1}^2 = 1$, implies
\begin{equation}
g_Y^2  \sum_i \left (\frac{c_i}{g'_i} \right)^2 =1 \, .
\end{equation}
Then, the condition 
\begin{equation}
P \equiv \frac{1}{g_Y^2} - \sum_i \left(\frac{c_i}{g'_i} \right)^2  =0 
\label{perp}
\end{equation}
encodes the orthogonality of the mixing matrix connecting the fields
coupled to the flavor charges $Q_1$, $Q_2$, $Q_3,\, \dots$ and the
  fields rotated, so that one of them, $Y$, couples to the hypercharge
  $Q_Y$. Therefore, for orthogonal charges,  as the couplings  run 
with energy, {\em the condition $P=0$ needs to stay intact~\cite{Anchordoqui:2011eg}.} 

A very important point is that the couplings that are running are those of the $U(1)$ fields; hence the $\beta$ functions receive contributions from fermions and scalars, but not from gauge bosons. As a consequence, if we start with a set of couplings at a high mass scale $\Lambda$ satisfying $P=0$, this condition will be mantained at one loop as the couplings run down to lower energies ($Q$).  The one loop correction to the various couplings are 
\begin{equation} \frac{1}{\alpha_Y(Q)} = \frac{1}{\alpha_Y (\Lambda)} - \frac{b_Y}{2\pi} \, \ln(Q/\Lambda) \,, \end{equation} \begin{equation} \frac{1}{\alpha_i(Q)} = \frac{1}{\alpha_i (\Lambda)} - \frac{b_i}{2\pi} \, \ln(Q/\Lambda) \,, \label{RGbi} \end{equation} where \begin{equation} b_Y = \frac{2}{3} \, {\rm Tr} \, Q_{Y,f}^2 \, + \frac{1}{3} \, {\rm Tr} \, Q_{Y,s}^2, \end{equation}
and 
\begin{equation} b_i = \frac{2}{3} \, {\rm Tr} \, Q_{i,f}^2 \, + \frac{1}{3} \, {\rm Tr} \, Q_{i,s}^2, \end{equation}
with $f$ and $s$ indicating contribution from fermion and scalar loops, respectively.

Recall that the charges are orthogonal, $\sum_sQ_{i,s}Q_{k,s}=\sum_fQ_{i,f}Q_{k,f}=0$ for $i\neq k$. Then Eq.(\ref{normalization}) implies
\begin{equation}
\sum_sQ_{Y,s}^2=\sum_i c_i^2\sum_sQ_{i,s}^2  \quad \quad {\rm and} \quad \quad \sum_fQ_{Y,f}^2=\sum_i c_i^2\sum_fQ_{i,f}^2 \,,\label{ysum}
\end{equation}
hence
\begin{equation} b_Y = \sum_i c_i^2 b_i\ . \end{equation}  
 On the other, the RG-induced change of $P$ defined in Eq.(\ref{perp}) reads
\begin{eqnarray}
  \Delta P & = & \Delta \left(\frac{1}{\alpha_Y}\right) - \sum_i c_i^2 \, \Delta \left(\frac{1}{\alpha_i}\right) \nonumber \\
             & = & \frac{1}{2\pi} \left(b_Y - \sum_i c_i^2 \, b_i \right) \, \ln(Q/\Lambda)  \, .
\label{deltap}
\end{eqnarray}
Thus, $P=0$ stays valid to one loop if the charges are orthogonal~\cite{Anchordoqui:2011eg}. 

Should the charges not be orthogonal, it is instructive to write Eq.~(\ref{3}) as
$\mathbb{ V} \, .\, \mathbb{Q} = 0$, where 
\begin{equation}
V_i = O_{i1} - \frac{g_Y \, c_i}{g'_i} \, .
\end{equation} 
 Certainly  $V_i =0$ still holds as a possible solution. But as the charges do not form a mutually orthogonal basis, one can ask whether other solutions exist. This will be the case if, for non-zero $\mathbb{V}$, 
\begin{equation}
\sum_i V_i \, Q_i^\alpha = 0
\label{xavi}
\end{equation}
for each $\alpha$, where $Q_i^\alpha$ is the $U(1)$ charge of the particle $\alpha$. In the $U(3) \times U(2) \times U(1)$ gauge group of~\cite{Antoniadis:2000ena}, the right-handed electron is charged only with respect to one of the abelian groups. From  (\ref{xavi}), this sets one of the $V$'s (say $V_1$) equal to zero. For $\alpha = Q_i,\,  U_i,\ D_i,\, L_i,\, E_i,\, N_i,\, H $, there remain at least 4 additional equations satisfied by the remaining components $V_2$ and $V_3$. The resulting overcompleteness leads to  $V_2 = V_3 = 0$. 

Although in most models the condition $P=0$ holds in spite of  the non-orthogonality of the $Q_i$'s, the RG equations controlling the running of the couplings lose their simplicity. In particular, since 
\begin{equation}
{\rm Tr} \, Q_Y^2 \neq \sum_i c_i^2 \, {\rm Tr} \, Q_i^2\,,
\end{equation}
the RG equations become coupled. In addition,  kinetic mixing is generated at one loop level even if it is absent initially~\cite{delAguila:1988jz}. Removal of the mixing term in order to restore canonical gauge kinetic energy requires an additional $O(3)$ rotation, greatly complicating the analysis.

Here, we are considering models where the underlying symmetry at high energies is $U(N)$ rather than $SU(N)$. Following~\cite{Antoniadis:2000ena} we normalize all $U(N)$ generators according to
\begin{equation}
\label{norm}
{\rm Tr}(T^{a}T^{b})={1\over 2}\delta^{ab}  \, ,
\end{equation}
and measure the corresponding $U(1)_N$ charges with respect to the coupling $g_N/\sqrt{2N}$, with $g_N$ the $SU(N)$ coupling constant. Hence, the fundamental representation of $SU(N)$ has $U(1)_N$ charge unity. Another important element of the RG analysis is that the $U(1)$ couplings ($g'_1, g'_2, g'_3$)  run different from the non-abelian $SU(3)$ ($g_3$) and $SU(2)$ ($g_2$). This implies that the previous relation for normalization of abelian and non-abelian coupling constants, $g_N' = g_N/\sqrt{2N},$ holds only at the scale of $U(N)$ unification~\cite{Anchordoqui:2011eg}. The SM chiral fermion charges  in Table~\ref{t:spectrum} are not orthogonal as given (${\rm Tr} \, Q_{1L} \, Q_{1R} \neq 0,$).  Orthogonality can be completed by including a right-handed neutrino. 

An obvious question is whether each of the fields on the rotated basis couples to a single charge $\bar Q_i$. Let
\begin{equation}
{\mathscr L}  = \mathbb{X}^T \mathbb{G} \mathbb{Q} \,,
\label{milito}
\end{equation}
be the Lagrangian in the $1,\ 2,\ 3,\ \dots$ basis, with $X_\mu^i$ and  $Q_i$ vectors and $\mathbb{G}$ a diagonal matrix in $N$-dimensional 'flavor' space.  Now rotate to new orthogonal basis ($\bar{\mathbb{Q}}$) for $\mathbb{Q}$
\begin{equation}
\mathbb{Q} = \mathbb{R} \bar {\mathbb{Q}} \,;
\end{equation}
(\ref{milito}) becomes
\begin{equation}
{\mathscr L}  = \mathbb{X}^T \mathbb{G} \mathbb{R} \bar{\mathbb{Q}} \, .
\end{equation}
As it stands, each $X_\mu^i$ does not couple to a unique charge $\bar Q_i$; hence we rotate $\mathbb{X}$,
\begin{equation}
\mathbb{X} = \mathbb{O} \bar{\mathbb {Y}},
\end{equation}
to obtain
\begin{equation}
{\mathscr L} = \bar{\mathbb{Y}}^T \mathbb{O}^T \mathbb{G} \mathbb{R} \bar{\mathbb{Q}} \, .
\end{equation}
We wish to see if, for  given $\mathbb{O}$ and $\mathbb{G}$, we can find an $\mathbb{R}$ so that 
\begin{equation}
\mathbb{O}^T \mathbb{G} \mathbb{R} = \bar{\mathbb{G}} \, ({\rm diagonal})
\label{piletaya}
\end{equation}
This allows each $\bar Y_\mu^i$ to couple to a unique charge $\bar Q_i$ with strength $\bar g_i$. To see the problem with this, we rewrite (\ref{piletaya}) in terms of components
\begin{equation}
(O^T)_{ij} g_j R_{jk} = \bar g_i \delta_{ik} \, ;
\label{MORSANADA}
\end{equation}
for $i \neq k$, (\ref{MORSANADA}) leads to
\begin{equation}
(O^T)_{ij} \, g_j\, R_{jk} = 0 \ .
\label{smores}
\end{equation}
In general, in Eq.~(\ref{smores}) there are $N(N-1)$ equations, but only $N (N-1)/2$ independent $O_{ij}$ generators in $SO(N)$; therefore the system is overdetermined~\cite{Haim}. Of course, if $\mathbb{G} = g \mathbb{I}$, the equation becomes
\begin{equation}
\mathbb{O}^T \mathbb{R} = \mathbb{I},
\end{equation}
and so $\mathbb{O} =  \mathbb{R}$. 

We illustrate with the case $N=2$;
let
\begin{eqnarray}
\mathbb{R} & = & \left(\begin{array}{cc} C_\varphi & S_\varphi \\ -S_\varphi & C_\varphi \end{array} \right) \nonumber \\
\mathbb{G} & = & \left(\begin{array}{cc} g'_1 & 0 \\ 0 & g'_3 \end{array} \right)  \\
\mathbb{O} & = & \left(\begin{array}{cc} C_\vartheta & S_\vartheta \\ -S_\vartheta & C_\vartheta \end{array} \right) \nonumber
\end{eqnarray}
then
\begin{equation}
\mathbb{O} \mathbb{G} \mathbb{R}  =  \left( \begin{array}{cc} g'_1 C_\vartheta C_\varphi + g'_3 S_\vartheta S_\varphi & g'_1 C_\vartheta S_\varphi - g'_3 S_\vartheta C_\varphi \\
g'_1 S_\vartheta C_\varphi + g'_3 C_\vartheta S_\varphi & g'_1 S_\vartheta S_\varphi - g'_3 C_\vartheta C_\varphi \end{array}\right) 
 =  \left(\begin{array}{cc} \bar{g}'_1 & 0 \\ 0 & \bar{g}'_3 \end{array} \right)  \, .
\end{equation} 
From the off diagonal terms we obtain
\begin{eqnarray}
g'_1 C_\vartheta S_\varphi - g'_2 S_\vartheta C_\varphi = 0 & \Rightarrow & \tan \vartheta = \frac{g'_1}{g'_2} \tan \varphi \nonumber \\
g'_1 S_\vartheta C_\varphi - g'_2 C_\vartheta S_\varphi = 0 & \Rightarrow & \tan \vartheta = \frac{g'_2}{g'_1} \tan \varphi \nonumber 
\end{eqnarray}
which implies that $g'_1 = g'_2 = g$, or equivalently that $\mathbb{G}$ is a multiple of the unit matrix. Next, we consider the diagonal elements using $g'_1 = g'_2$ to obtain
\begin{eqnarray}
\cos (\vartheta - \varphi) = 0 & \Rightarrow & \vartheta = \varphi
\end{eqnarray}
Note that the matrix $\mathbb{R}$ has one independent variable, and there are two independent homogeneous equations.

Any vector boson $Y_\mu'$, orthogonal to the hypercharge, must grow a mass $M'$ in order to avoid long range forces  between baryons other than gravity and Coulomb forces. The anomalous mass growth allows the survival of global baryon number conservation, preventing fast proton decay~\cite{Ghilencea:2002da}. It is this that we now turn to sutdy.

\section{Premises of the anomalous sector}
Outside of the Higgs couplings, the relevant parts of the Lagrangian
are the gauge couplings generated by the $U(1)$ covariant derivatives
acting on the matter fields, and the (mass)$^2$ matrix of the
anomalous sector
\begin{equation}
{\mathscr L}  = \mathbb{Q}^T\  \mathbb{G}\ \mathbb{X}  + \tfrac{1}{2} \mathbb{X}^T \, \mathbb{M}^2 \,\mathbb{X} \,,
\end{equation}
where $X_\mu^i$ are the three $U(1)$ gauge fields in the D-brane basis ($B_\mu,\, C_\mu,\, \tilde B_\mu$), $\mathbb{G}$ is a diagonal coupling matrix 
$(g'_1, g'_3, g'_4)$, and $\mathbb{Q}$ are the 3 charge matrices.

Again, perform a rotation $\mathbb{X}= \mathbb{O} \, \mathbb{Y}$ and
require that one of the $\mathbb{Y}$'s (say $Y_\mu$) couple to
hypercharge. We then obtain the constraint on the first column of
$\mathbb{O}$ given in (\ref{tooslow}).  However, there is now an
additional constraint: {\em the field $Y_\mu$ is an eigenstate of
  $\mathbb{M}^2$ with zero eigenvalue.} Under the $\mathbb{O}$
rotation, the mass term becomes
\begin{equation}
\tfrac{1}{2} \mathbb{X}^T \mathbb{M}^2 \mathbb{X} = \tfrac{1}{2} \mathbb{Y}^T\ \overline{\mathbb{M}^2}\ \ \mathbb{Y} \,,
\end{equation}
with $\overline{\mathbb{M}^2} = \mathbb{O}^T\ \mathbb{M}^2 \ \mathbb{O} .$ We know that at least $Y_\mu$ is an eigenstate with eigenvalue zero. We also know that Poincare invariance requires the complete diagonalization of the mass matrix in order to deal with observables. However, further similarity transformations will undo the coupling of the zero eigenstate to hypercharge. There seems no way of eventually fulfilling all these conditions except to require that the same $\mathbb{O}$ which rotates to couple $Y_\mu$ to hypercharge simultaneously diagonalizes $\mathbb{M}^2$ so that
\begin{equation}
\overline{\mathbb{M}^2} = {\rm diag} (0, M'^2, M''^2) \, .
\end{equation}
This implies that the original $\mathbb{M}^2$ in the flavor basis is given by
\begin{equation}
\mathbb{M}^2 = \mathbb{O}\ {\rm diag} (0, M'^2, M''^2) \mathbb{O}^T \,,
\end{equation}
which results in the following baroque matrix: 
\begin{equation}
\mathbb{M}^2 = \left( \begin{array}{ccc} a & b & c \\ b& d & e \\ c & e & f \end{array} \right) \,, 
\end{equation}
where
\begin{eqnarray}
  a & = &  M'^2 (C_\psi S_\theta S_\phi - C_\phi S_\psi)^2 + M''^2 (C_\phi C_\psi S_\theta + S_\phi S_\psi)^2 \,, \nonumber \\
 b & = & ({M'}^2-{M''}^2) C_\phi C_{2\psi} S_\theta S_\phi + C^2_\phi C_\psi (-{M'}^2 +{M''}^2 S^2_\theta) S_\psi + C_\psi 
(-{M''}^2 + {M'}^2 S_\theta^2) S_\phi^2 S_\psi  \, , \nonumber \\
c & = &  C_\theta [{M''}^2 C^2_\phi C_\psi S_\theta + {M'}^2 C_\psi S_\theta S^2_\phi - ({M'}^2-{M''}^2) C_\phi S_\phi S_\psi] \,, \nonumber\\
d & = & {M''}^2 (C_\psi S_\phi - C_\phi S_\theta S_\psi)^2 + {M'}^2 (C_\phi C_\psi + S_\theta S_\phi S_\psi)^2 \, , \nonumber \\
e & = & C_\theta [({M'}^2 - {M''}^2) C_\phi C_\psi S_\phi + {M''}^2 C_\phi^2 S_\theta S_\psi + {M'}^2 S_\theta S_\phi^2 S_\psi ] \, , \nonumber \\
 f & = & C_\theta^2 ({M''}^2 C^2_\phi + {M'}^2 S^2_\phi ) \, .
\label{biguglyMF}
\end{eqnarray}

We turn now to discuss the phenomenological aspects of anomalous $U(1)$ gauge bosons related to experimental searches for new physics at the Tevatron and at the CERN's Large Hadron Collider (LHC).

\section{Search for New Gauge Bosons at  Hadron Colliders}

Taken at face value, the disparity between
CDF~\cite{Aaltonen:2011mk,Punzi} and  D\O~\cite{Abazov:2011af} results
insinuates a commodious uncertainty as to whether there is an excess
of events in the dijet system invariant mass distribution of the
associated production of a $W$ boson with 2 jets (hereafter $Wjj$
production). The $M_{jj}$ excess showed up in $4.3~{\rm fb}^{-1}$ of
integrated luminosity collected with the CDF detector as a broad bump
between about 120 and 160~GeV~\cite{Aaltonen:2011mk}. The CDF
Collaboration fitted the excess (hundreds of events in the $\ell jj +
\met$ channel) to a Gaussian and estimated its production cross
section times the dijet branching ratio to be $4~{\rm pb}$. This is
roughly 300 times the SM Higgs rate $\sigma (p\bar p
\to WH) \times {\rm BR} (H \to b \bar b)$. For a search window of $120
- 200~{\rm GeV}$, the excess significance above SM background
(including systematics uncertainties) has been reported to be
$3.2\sigma$~\cite{Aaltonen:2011mk}. Recently, CDF has included an
additional $3~{\rm fb}^{-1}$ to their data sample, for a total of
$7.3~{\rm fb}^{-1}$, and the statistical significance has grown to
$\sim 4.8\sigma$ ($\sim 4.1\sigma$ including
systematics)~\cite{Punzi}. More recently, the  D\O\ Collaboration
released an analysis (which closely follows the CDF analysis) of their
$Wjj$ data finding ``no evidence for anomalous resonant dijet
production''~\cite{Abazov:2011af}. Using an integrated luminosity of
$4.3~{\rm fb^{-1}}$ they set a 95\% CL upper limit of $1.9~{\rm pb}$
on a resonant $Wjj$ production cross section.

Although various explanations have been proposed for the CDF anomaly~\cite{Eichten:2011sh},
perhaps the simplest is the introduction of a new leptophobic $Z'$ gauge boson~\cite{Buckley:2011vc}. 
The suppressed coupling to leptons (or more specifically, to electrons
and muons) is required to evade the strong constraints of the Tevatron
$Z'$ searches in the dilepton mode~\cite{Acosta:2005ij} and LEP-II
measurements of $e^+ e^- \to e^+ e^-$ above the
$Z$-pole~\cite{Barate:1999qx}. In complying with the precision
demanded of our phenomenological approach it would be sufficient to
consider a 1\% branching fraction to leptons as consistent with the
experimental bound. This approximation is within a factor of a few of
{\em model independent} published experimental bounds. In addition,
the mixing of the $Z'$ with the SM $Z$ boson should be extremely small
to be compatible with precision measurements at the $Z$-pole by the
LEP experiments~\cite{Umeda:1998nq}.

All existing dijet-mass searches via direct production at the Tevatron are limited to $M_{jj} > 200~{\rm GeV}$~\cite{Abe:1993kb} and therefore cannot constrain the existence of a $Z'$ with $M_{Z'} \simeq 150~{\rm GeV}$. The strongest constraint on a light leptophobic $Z'$ comes from the dijet search by the UA2 Collaboration, which has placed a 90\% CL upper bound on $\sigma (p\bar p \to Z') \times {\rm BR}(Z' \to jj)$ in this energy range~\cite{Alitti:1990kw}.  A comprehensive model independent analysis incorporating Tevatron and UA2 data to constrain the $Z'$ parameters for predictive purposes at the LHC was recently presented~\cite{Hewett:2011nb}.\footnote{Other phenomenological restrictions on $Z'$-gauge bosons were recently discussed in~\cite{Williams:2011qb}.}  As of today the ATLAS and CMS experiments are not sensitive to the $Wjj$ signal~\cite{Eichten:2011xd}. However,  LHC will eventually weigh in on this issue: if new physics is responsible for the CDF anomaly, an excess in  $\ell jj +
\met$  should become statistically significant in ATLAS and CMS by the end of the year~\cite{Buckley:2011hi}.

As usual, the  $U(1)$ gauge interactions arise through the covariant derivative 
\begin{equation}\el{covderi2} \CD_\mu = \p_\mu - i g'_3 \, C_\mu \, Q_3 -i g'_4 \, \tilde B_\mu \, Q_{1L} -i g'_1 \, B_\mu \, Q_{1R} \, ,
\end{equation}
where $g'_1$, $g'_3$, and $g'_4$ are the  gauge coupling constants. 
The fields $C_\mu, \tilde B_\mu, B_\mu$ are related
to $Y_\mu, Y_\mu{}'$ and $Y_\mu{}''$ by the rotation matrix, 
\begin{equation}
\mathbb{O} = \left(
\begin{array}{ccc}
 C_\theta C_\psi  & -C_\phi S_\psi + S_\phi S_\theta C_\psi  & S_\phi
S_\psi +  C_\phi S_\theta C_\psi  \\
 C_\theta S_\psi  & C_\phi C_\psi +  S_\phi S_\theta S_\psi  & - S_\phi
C_\psi + C_\phi S_\theta S_\psi  \\
 - S_\theta  & S_\phi C_\theta  & C_\phi C_\theta
\end{array}
\right) \,,
\end{equation}
with Euler angles $\theta$, $\psi,$ and $\phi$. Equation~(\ref{covderi2}) can be rewritten in terms of $Y_\mu$, $Y'_\mu$, and
$Y''_\mu$ as follows
\begin{eqnarray}
\CD_\mu & = & \partial_\mu -i Y_\mu \left(-S_\theta g'_1 Q_{1R} + C_\theta S_\psi  g'_4  Q_{1L} +  C_\theta C_\psi g'_3 Q_3 \right) \nonumber \\
 & - & i Y'_\mu \left[ C_\theta S_\phi  g'_1 Q_{1R} +\left( C_\phi C_\psi + S_\theta S_\phi S_\psi \right)  g'_4 Q_{1L} +  (C_\psi S_\theta S_\phi - C_\phi S_\psi) g'_3 Q_3 \right] \label{linda} \\
& - & i Y''_\mu \left[ C_\theta C_\phi g'_1 Q_{1R} +  \left(-C_\psi S_\phi + C_\phi S_\theta S_\psi \right)  g'_4  Q_{1L} + \left( C_\phi C_\psi S_\theta + S_\phi S_\psi\right) g'_3 Q_3 \right]   \, .  \nonumber
\label{suraS}
\end{eqnarray}
Now, by demanding that $Y_\mu$ has the
hypercharge $Q_Y$ given in Eq.~\er{hyperchargeY}  we  fix the first column of the rotation matrix $\mathbb{O}$
\begin{equation}
\bay{c} C_\mu \\ \tilde B_\mu \\ B_\mu
\eay = \left(
\begin{array}{lr}
  Y_\mu \, c_3g_Y /g'_3& \dots \\
  Y_\mu \, c_4 g_Y/g'_4 & \dots\\
   Y_\mu \, c_1g_Y/g'_1 & \dots
\end{array}
\right) \, ,
\label{norby}
\end{equation}
and we determine the value of the two associated Euler angles
\begin{equation}
\theta = {\rm -arcsin} [c_1 g_Y/g'_1]
\label{theta}
\end{equation}
and
\begin{equation}
\psi = {\rm arcsin}  [c_4 g_Y/ (g'_4 \, C_\theta)] \, .
\label{psi}
\end{equation}
The couplings $g'_1$ and $g'_4$ are related through the orthogonality condition (\ref{perp}),
\begin{equation}
 \left(\frac{c_4}{ g' _4} \right)^2  = \frac{1}{g_Y^2} - \left(\frac{c_3}{g'_3} \right)^2 - \left(\frac{c_1}{g'_1}\right)^2  \, ,
\end{equation}
with $g'_3$ fixed by the relation for $U(N)$ unification: $g'_3  = \sqrt{6} \, g_3$. In what follows, we take $5~{\rm TeV}$ as a reference point for running down to 150~GeV the $g'_3$ coupling using (\ref{RGbi}), that is {\em ignoring mass threshold effects}. This yields $g'_3 = 0.383$. We have checked that the running of the $g'_3$ coupling does not change significantly by varying the scale of $U(N)$ unification between $3~{\rm TeV}$ and $10~{\rm TeV}.$

The phenomenological analysis thus far has been formulated in terms
of the  mass-diagonal basis set of gauge fields $(Y,Y',Y'')$. As a
result of the electroweak phase transition, the coupling of this set
with the Higgses will induce mixing, resulting in a new mass-diagonal basis set
$(Z,Z',Z'')$. It will suffice to analyze only the $2\times 2$ system
$(Y,Y')$ to see that the effects of this mixing are totally
negligible. We consider simplified zeroth and first order (mass)$^2$
matrices
\begin{equation*}
 (M^2)^{(0)}\ =\ \left (\begin{array}{cc} 0 & 0 \\ 0 & M'^2 \end{array} \right) \quad
(M^2)^{(1)}\ = \ \left (\begin{array}{cc} \overline{M}_Z^2 & \epsilon \\ \epsilon & m'^2 \end{array} \right)
\end{equation*}
where $M'$ is the mass of the $Y'$ gauge field, $\overline{M}_Z = \sqrt{g_2^2 + g_Y^2} \ v/2$ is the usual tree level formula for the mass  of the $Z$ particle in the electroweak theory (before mixing), $g_2 \simeq 0.651$ is the electroweak coupling constant, $v$ is the vacuum expectation value of the Higgs field, 
$g_Y \simeq 0.357$, and $\epsilon, m'^2$ are of ${\cal O}(\overline{M}_Z^2)$.

Standard Rayleigh-Schrodinger perturbation theory then provides the (mass)$^2$ (to second order in
$\overline{M}_Z^2$) and
wave functions (to first order)  of the mass-diagonal eigenfields $(Z,Z')$ corresponding to $(Y,Y')$
\begin{equation}
M_Z^2 = \overline{M}_Z^2 - \left(\frac{\epsilon^2}{M'^2}\right) \,, \quad 
M_{Z'}^2 = M'^2 + m'^2 + \left(\frac{\epsilon^2}{M'^2}\right) \,,
\label{seag1}
\end{equation}
and
\begin{equation}
Z = Y - \left( \frac{\epsilon}{M'^2} \right)  \ \ Y' \,, \quad 
Z' = Y' +  \left(\frac{\epsilon}{M'^2} \right)\ \ Y  \, .
\label{seag2}
\end{equation}
From Eqs.~(\ref{seag1}) and (\ref{seag2}) the shift in the mass of the $Z$ is given by $\delta M_Z^2 = (\epsilon/M')^2,$ so that
$\epsilon = M' \sqrt{2 M_Z \delta M_Z}.$ The admixture of $Y$ in the mass-diagonal field $Z'$ is
\begin{equation}
 \theta = \frac{\epsilon}{M'^2}  = \frac{M_Z}{M'} \;\;\sqrt{\frac{2 \delta M_Z} {M_Z}} \simeq 0.004 \, .
\end{equation}
Since all effects go as $\theta^2\simeq 1.6\times 10^{-5}$, all further discussion will be, with negligible error, in terms of $Z'$.  By the same token, the admixture of $Y'$ in the eigenfield $Z$ is negligible, so that the discussion henceforth will reflect $Z \simeq Y$ and $\overline{M}_Z^2 \simeq M_Z^2$.

\begin{figure}[t]
\vspace*{1.0cm}
\[
\phantom{XXXXXXX}
\vcenter{
\hbox{
  \begin{picture}(0,0)(0,0)
\SetScale{1.5}
  \SetWidth{.3}
\ArrowLine(-45,20)(-25,20)
\Photon(-25,20)(-5,20){2}{6}
\ArrowLine(-45,-20)(-25,-20)
\Photon(-5,-20)(-25,-20){2}{6}
\ArrowLine(-25,20)(-25,-20)
\ArrowLine(-5,-20)(5,-13)
\ArrowLine(-5,-20)(5,-27)
\ArrowLine(-5,20)(5,27)
\ArrowLine(-5,20)(5,13)
\Text(-70,20)[cb]{{\footnotesize $q_1$}}
\Text(-70,-25)[cb]{{\footnotesize $\bar q_2$}}
\Text(-21,-45)[cb]{{\footnotesize $Z'$}}
\Text(-19,35)[cb]{{\footnotesize $W$}}
\Text(12,-22)[cb]{{\footnotesize $q$}}
\Text(12,-45)[cb]{{\footnotesize $\bar q$}}
\Text(12,39)[cb]{{\footnotesize $\nu$}}
\Text(-47,2)[cb]{{\footnotesize $q_2$}}
\Text(12,14)[cb]{{\footnotesize $\ell$}}
\end{picture}}
}
\]
\vspace*{.6cm}
\caption[]{Feynman diagram for $q\bar q \to WZ' \to \nu \ell jj$.}
\label{feynman}
\end{figure}
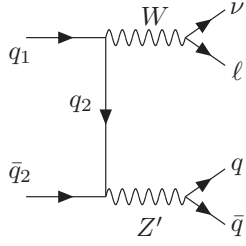

The $f \bar f Z'$ Lagrangian is of the form
\begin{eqnarray}
{\cal L} & = & \frac{1}{2}   \sqrt{g_Y^2 + g_2^2} \ \sum_f \bigg(\epsilon_{f_L} \bar \psi_{f_L} \gamma^\mu \psi_{f_L} +   \epsilon_{f_R} \bar \psi_{f_R} \gamma^\mu \psi_{f_R} \bigg) \, Z'_\mu \, \nonumber \\
& = & \sum_f \bigg((g_{Y'}Q_{Y'})_{f_L} \, \bar \psi_{f_L} \gamma^\mu \psi_{f_L} +  (g_{Y'}Q_{Y'})_{f_R} \bar \psi_{f_R} \gamma^\mu \psi_{f_R} \bigg) \, Z'_\mu \,
\label{lagrangian}
\end{eqnarray}
where each $\psi_{f_{L \, (R)}}$ is a fermion field  with the corresponding $\gamma^\mu$  matrices of the Dirac algebra, and $\epsilon_{f_L,f_R} = v_q \pm a_q$, with $v_q$ and $a_q$ the vector and axial couplings respectively.  The (pre-cut) $Wjj$ production rate  at the Tevatron $\sqrt{s} = 1.96~{\rm pb}$, for arbitrary couplings and $M_{Z'} \simeq 150~{\rm GeV}$, is found to be~\cite{Hewett:2011nb}
\begin{equation}
\sigma (p\bar p \to WZ') \times {\rm BR} (Z' \to jj)  \simeq  \left[0.719 \left(\epsilon_{u_L}^2 + \epsilon_{d_L}^2\right) + 5.083 \, \epsilon_{u_L}  \, \epsilon_{d_L} \right] \times \Gamma (\phi, g'_1)_{Z'\to q \bar q}~{\rm pb}  \, ,
\label{wjj-production}
\end{equation}
where $\Gamma (\phi, g'_1)_{Z'\to q \bar q}$ is the hadronic branching fraction. The presence of a $W$ in the process shown in Fig.~\ref{feynman} restricts the contribution of the quarks to be purely left-handed. Since $\epsilon_{u_L} = \epsilon_{d_L}$ and the required branching to quarks is above about 99\%  (after selection cuts are accounted for) the coupling strength  $\epsilon^2_{q_L}$ is fixed by the Wjj production rate. Below, we avoid reference to specific experimental selection cuts and present results for a generous range of possibilitites consistent with existing data. 

\begin{figure}[tbp]
\postscript{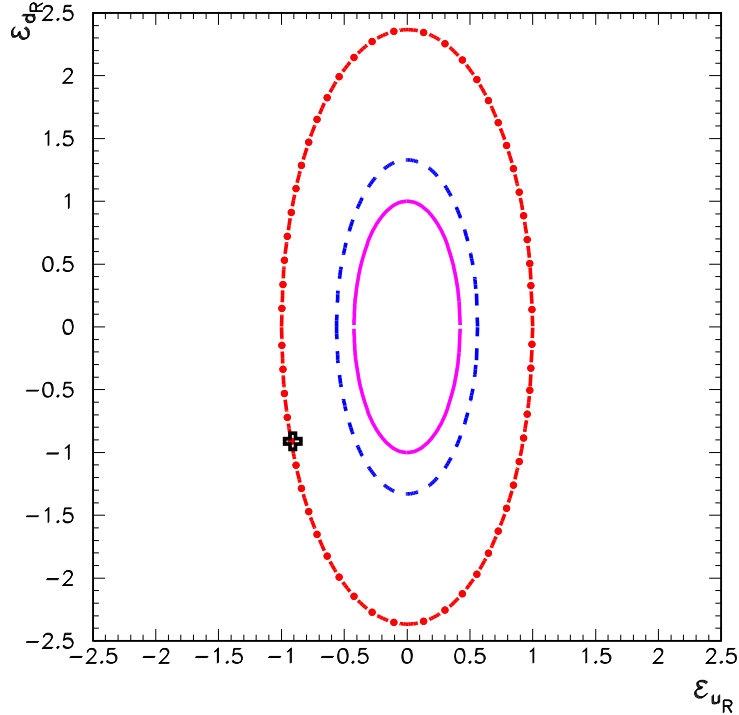}{0.6}
\caption{The ellipses show  the  values of  $\epsilon_{u_R}$ and $\epsilon_{d_R}$ that saturate  the UA2 and  D\O\ limits on direct $Z' \to jj$ and $Wjj$ production, respectively. The solid ellipse is based on the assumption that experimental selection cuts will cause negligible reduction in event rates, the dashed ellipse corresponds to a reduction in event rates by 50\%, and the dot-dashed ellipse corresponds to a 66\% reduction in  D\O\ event rates and 70\% reduction in UA2 event rates. The cross indicates the best eyeball fit that simultaneously ensures  small $Z-Z'$ mixing and $\Gamma_{Z' \to e^+ e^-}/\Gamma_{Z'\to q \bar q} \alt 1\%$.} 
\label{HR}
\end{figure}

The dijet  production rate at the UA2 $\sqrt{s} = 630~{\rm GeV}$ can be parametrized as follows~\cite{Hewett:2011nb}
\begin{equation}
\sigma (p\bar p \to Z' ) \times {\rm BR} (Z'\to jj) \simeq \tfrac{1}{2} \big[ 773  ( \epsilon_{u_L}^2 + \epsilon_{u_R} ^2) + 138  ( \epsilon_{d_L}^2 + \epsilon_{d_R} ^2)\big] \times \Gamma(\phi,g'_1)_{Z' \to q  \bar q}~{\rm pb} \, .
\label{zprime-production}
\end{equation}
(Our numerical calculation~\cite{Anchordoqui:2011eg} using  CTEQ6~\cite{Pumplin:2002vw} agrees within 5\% with the result of~\cite{Hewett:2011nb}.)  The maximum allowed value of the $\epsilon_{u_R}$ and $\epsilon_{d_R}$ couplings consistent with the UA2 upper limit are shown in Fig.~\ref{HR}. The dilepton production rate at UA2 energies is given by
\begin{equation}
\sigma (p\bar p \to Z' ) \times {\rm BR} (Z'\to \ell \bar \ell) \simeq \tfrac{1}{2} \big[ 773  ( \epsilon_{u_L}^2 + \epsilon_{u_R} ^2) + 138  ( \epsilon_{d_L}^2 + \epsilon_{d_R} ^2)\big] \times \Gamma (\phi, g'_1)_{Z' \to \ell  \bar \ell}~{\rm pb} \, ,
\label{zprimeleptons}
\end{equation}
where $\Gamma(\phi, g'_1)_{Z' \to \ell  \bar \ell}$ is the leptonic branching fraction. From (\ref{linda}) and (\ref{lagrangian}) we obtain the explicit form of the chiral couplings in terms of $\phi$ and $g'_1$
\begin{eqnarray}
\epsilon_{u_L} = \epsilon_{d_L} & = & \frac{2}{\sqrt{g_Y^2 + g_2^2}} \, (C_\psi S_\theta S_\psi - C_\phi S_\psi) g'_3 \,, \nonumber \\
\epsilon_{u_R} & = &- \frac{2}{\sqrt{g_Y^2 + g_2^2}} \, [C_\theta S_\phi g'_1 + (C_\psi S_\theta S_\psi - C_\phi S_\psi) g'_3] \,, \label{couplingphig}\\
\epsilon_{d_R} & = & \frac{2}{\sqrt{g_Y^2 + g_2^2}} \, [C_\theta S_\phi g'_1 - (C_\psi S_\theta S_\psi - C_\phi S_\psi) g'_3] \, . \nonumber
\end{eqnarray}

The second strong constraint on the model derives from the mixing of the $Z$ and the $Y'$ through their coupling to the Higgs doublet. The last two terms in the covariant derivative \begin{equation} \CD_\mu = \p_\mu - i \frac{1}{\sqrt{g_2^2 + g_Y^2}} Z_\mu (g_2^2 T^3 - g_Y^2 Q_Y) -i g_{Y'} Y_\mu{}' Q_{Y'} - i g_{Y''} Y_\mu{}'' Q_{Y''} , \end{equation} are conveniently written as \begin{equation} -i \frac {x_H}{ v} \overline M_Z Y_\mu{}' - i \frac{y_H} {v} \overline M_Z Y_\mu{}'' \end{equation} where \begin{equation} x_H = 1.9 \sqrt{{g'_1}^2 -0.032} \, S_\phi \,, \end{equation}
 \begin{equation}
y_{^H} =  1.9 \sqrt{{g'_1}^2 -0.032}
 \, C_\phi  \, ,
\end{equation}
and $T^3 = \sigma^3/2$.
The Higgs field  kinetic term $(D_\mu H)^\dagger (D_\mu H)$ together with the anomalous mass terms  ($-\frac{1}{2} M'^2 Y'_\mu Y'^\mu - \frac{1}{2} M''^2 Y''_\mu Y''^\mu$) yield the following mass square matrix\footnote{We note in passing that two
`supersymmetric' Higgses $H_u\equiv H$ and $H_d = H^\dagger$, with charges $Q_3 = Q_{1L} = 0$, $Q_{1R} = 1$, $Q_Y = 1/2$ and $Q_3 = Q_{1L} = 0$,
$Q_{1R} = -1$, $Q_Y = -1/2$, 
would also be sufficient to give masses to all
the chiral fermions.  Here, $\br H_u \ke = (^0_{v_u})$, $\br H_d \ke
= (^{v_d}_0),$ $v = \sqrt{v_u^2 + v_d^2}$, and  
$\tan \beta \equiv v_u/v_d$. It is easily seen that the corresponding mass square matrix is independent of $\tan \beta $~\cite{Anchordoqui:2011eg}.}  
\begin{equation}
\bay{ccc} \overline M_Z^2 & \overline M_Z^2 x_{H} & \overline M_Z^2 y_{H}  \\ \overline M_Z^2 x_{H}  & \overline
M_Z^2 x_{H}^2 + M'^2 & \overline M_Z^2
x_{H} y_{H}  \\
\overline M_Z^2 y_{H} & \overline M_Z^2 x_{H} y_{H} & \overline M_Z^2
y_{H}^2 + M''^2\eay \, .
\end{equation}

Next, taking $M_{Z'} = 150~{\rm GeV}$ we use the two degrees of
freedom of the model $(g'_1, \phi)$ to demand the shift of the $Z$
mass to lie within 1 standard deviation of the experimental value and
leptophobia. This occurs for $g'_1 = 0.2$, $\phi = 0.0028$ and
$M_{Z''} = 5~{\rm TeV}$, corresponding to a suppression $\Gamma_{Z''
  \to e^+ e^-}/\Gamma_{Z''\to q \bar q} \simeq 1\%$~\cite{Anchordoqui:2011eg}.  This also
corresponds to $\theta = -1.103$, $\psi = -1.227$, and $g'_4 =
0.42$. The $g_{Y'}Q_{Y'}$ and $g_{Y''} Q_{Y''}$ couplings to the
chiral fields are fixed and given in Table~\ref{case3}. The accompanying
values of $\epsilon_{u_R}$ and $\epsilon_{d_R}$ are shown in Fig.~\ref{HR}. Now,
substituting the above figures into (\ref{suraS}) we obtain the
projections over $Y,\ Y',\ Y''$
\begin{eqnarray}
Y & = & 1.8 \times 10^{-1}~Q_{1R} + 5.9 \times 10^{-2}~Q_3 - 1.8 \times 10^{-1}~Q_{1L} \nonumber \\                    
Y'& = & 2.5 \times 10^{-4}~Q_{1R} + 3.7 \times 10^{-1}~Q_3 + 1.4\times 10^{-1}~Q_{1L}  \\ 
Y'' & = & 9.0 \times 10^{-2}~Q_{1R} - 1.2 \times 10^{-1}~Q_3 + 3.5 \times 10^{-1}~Q_{1L} \nonumber  \, . 
\end{eqnarray}
Using Eq.~(\ref{bb-l}) it is straightforward to see that $Z'$ and $Z''$ become essentially $B$ and $B-L$, respectively.

The $Z'$ couplings to quarks leads to a large (pre-cut) $Wjj$ production ($\simeq 6~{\rm pb}$) at the Tevatron, and at $\sqrt{s} = 630~{\rm GeV}$, a direct (pre-cut) $Z' \to jj$ production ($\simeq 700~{\rm pb}$) in the region excluded by UA2 data. However, it is worthwhile to point out that the UA2 Collaboration performed their analysis in the early days of QCD jet studies. Their upper bound depends crucially on the quality of the Monte Carlo and detector simulation which are primitive by today's standard. They also use events with two exclusive jets, where jets were constructed using an infrared unsafe jet algorithm~\cite{Kunszt:1992tn}. In view of the considerable uncertainties associated with the UA2 analysis we remain skeptical of drawing negative conclusions. Instead we argue that our  model~\cite{Anchordoqui:2011eg}  could provide an explanation of the CDF anomaly if acceptance and pseudorapidity cuts reduce the $Wjj$ production rate by about 35\% - 66\% and the UA2 90\% CL bound is taken as an order-of-magnitude limit~\cite{UA2-comment}.

\begin{table}
\caption{Chiral couplings of $Y'$ and $Y''$ gauge bosons for $\phi = 0.0028$ and $g'_1 = 0.2$.}
\begin{tabular}{c|cc}
\hline
\hline
 Name &~~ $g_{Y'}Q_{Y'}$~ &~~$g_{Y''} Q_{Y''}$\\
\hline
~~$Q_i$~~ &  $ 0.368$ & $ -0.119$ \\[1mm]
~~$U_i$~~ & $0.368$ & $-0.028$ \\[1mm]
~~$D_i$~~ &  $0.368 $ &  $-0.209$ \\[1mm]
~~$L_i$~~ & $0.143$ & $\phantom{-} 0.143$\\[1mm]
~~$E_i$~~ & $0.142$ & $\phantom{-} 0.262$\\[1mm]
~~$N_i$~~ & $0.143$ & $\phantom{-} 0.443$ \\ [1mm]
\hline
\hline
\end{tabular}
\label{case3}
\end{table}

\begin{table}
\caption{Chiral couplings of $Y'$ and $Y''$ gauge bosons for $\phi = -0.0638$ and $g'_1 = 0.195$.}
\begin{tabular}{c|cc}
\hline
\hline
 Name &~~ $g_{Y'}Q_{Y'}$~ &~~$g_{Y''} Q_{Y''}$\\
\hline
~~$Q_i$~~ &  $0.370$ & $ -0.112$ \\[1mm]
~~$U_i$~~ & $0.365$ & $- 0.033$ \\[1mm]
~~$D_i$~~ &  $0.375 $ &  $- 0.190$ \\[1mm]
~~$L_i$~~ & $0.154$ & $\phantom{-} 0.154$\\[1mm]
~~$E_i$~~ & $0.159$ & $\phantom{-} 0.338$\\[1mm]
~~$N_i$~~ & $0.149$ & $\phantom{-}0.495$\\[1mm]
\hline
\hline
\end{tabular}
\label{case4}
\end{table}

\begin{figure}[tbp] \begin{minipage}[t]{0.49\textwidth} \postscript{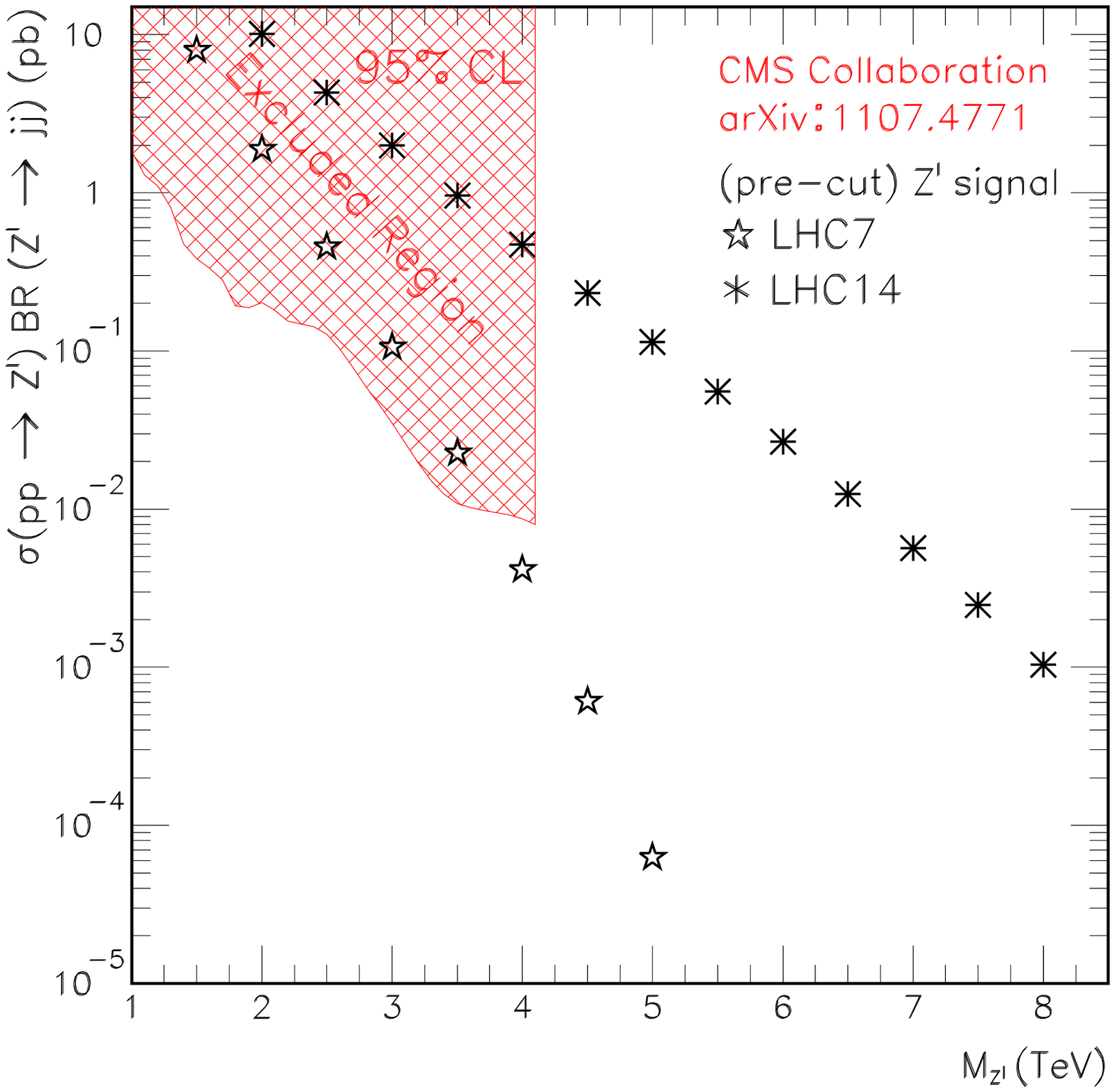}{0.99} \end{minipage} \hfill \begin{minipage}[t]{0.49\textwidth} \postscript{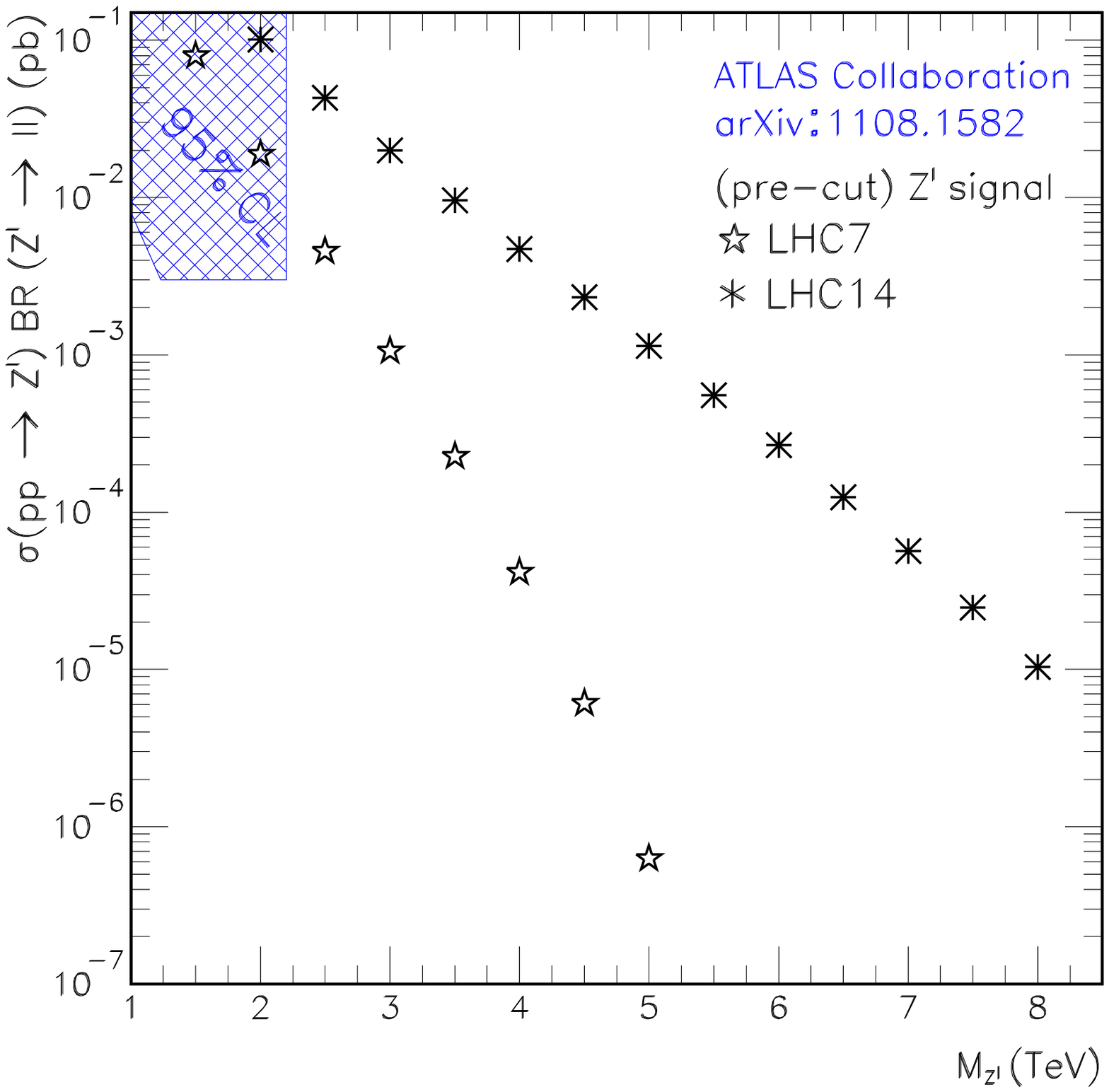}{0.99} \end{minipage} \caption{Comparison of the (pre-cut) total cross section for the production of $p p \to Z' \to jj$ (left) and $p p \to Z' \to \ell \ell$ (right) with the 95\% CL upper limits on the production of a gauge boson decaying into two jets (left) and two leptons (right), as reported by the CMS (corrected by acceptance)~\cite{Chatrchyan:2011ns} and ATLAS~\cite{Collaboration:2011dc} collaborations, respectively. We have taken $\phi = -0.0638$, $g'_1 = 0.195$. For isotropic decays (independently of the resonance), the acceptance for the CMS detector has been reporetd to be ${\cal A} \approx 0.6$~\cite{Chatrchyan:2011ns} . The predicted $Z'$ production rates are shown for $\sqrt{s} = 7~{\rm TeV}$ and $\sqrt{s} = 14~{\rm TeV}$.}  
\label{LHC-Zprime} \end{figure}

Since the CDF signal is in dispute, it is of interest to study the predictions of the model for a leptophobic $Z'$ at energies not obtainable at the Tevatron, but within the range of the LHC. The ATLAS Collaboration has searched for narrow resonances in the invariant mass spectrum of dimuon and dielectron final states in event samples corresponding to an integrated luminosity of $1.21~{\rm
  fb}^{-1}$ and $1.08~{\rm fb}^{-1}$, respectively~\cite{Collaboration:2011dc}. The spectra are consistent with SM expectations and thus a lower mass limit of $1.83~{\rm TeV}$ on the sequential SM $Z'$ has bee set.\footnote{In the sequential SM the $Z'$ has the same couplings to fermions as the $Z$ boson.}  Therefore, for $M_{Z'} \geq 1~{\rm TeV}$, we scan the $g'_1-\phi$ paramenter space demanding the shift of the $Z$ mass to lie within 1 standard deviation of the experimental value and small $(\alt 1\%)$ branching to leptons. We find that for $g'_1 = 0.195$, $\phi = -0.0638$, and $M_{Z''} \geq 2 M_{Z'}$, the ratio $\Gamma_{Z' \to e^+ e^-}/\Gamma_{Z'\to q \bar q} \alt 1\%$~\cite{Anchordoqui:2011eg}. The chiral couplings to the $Z'$ and $Z''$ gauge bosons for these fiducial values are given in Table~\ref{case4}. Again, we see that $Z'$ and $Z''$ are essentially $B$ and $B-L$.

The decay width of $Z' \to f\bar f$ is given by~\cite{Barger:1996kr}
\begin{equation}
\Gamma( Z' \to f \bar f) = \frac{G_F M_Z^2}{6 \pi \sqrt{2}}  N_C C(M_{Z'}^2) M_{Z'} \sqrt{1 -4x} \left[v_f^2 (1+2x) + a_f^2 (1-4x) \right] \, ,
\end{equation}
where $G_F$ is the Fermi coupling constant, $C(M_{Z'}^2) = 1 +
\alpha_s/\pi + 1.409 (\alpha_s/\pi)^2 - 12.77 (\alpha_s/\pi)^3$,
$\alpha_s = \alpha_s(M_{Z'})$ is the strong coupling constant at the
scale $M_{Z'}$, $x = m_f^2/M_{Z'}^2$, $v_f$ and $a_f$ are the vector
and axial couplings, and $N_C =3$ or 1 if $f$ is a quark or a lepton,
respectively. Using the fiducial values of $g'_1$ and $\phi$ fitted in
Table~\ref{case3}, for $M_{Z'} = 1~{\rm TeV}$, we obtain $\Gamma =
60.9~{\rm TeV}$. Hence, to compare our predictions (at the parton
level) with LHC experimental searches in dilepton and dijets it is
sufficient to consider the production cross section in the narrow $Z'$
width approximation,
\begin{equation}
\hat \sigma (q \bar q \to Z')  =   K \frac{2 \pi}{3} \, \frac{G_F \, M_Z^2}{\sqrt{2}}  \left[v_q^2 (\phi, g'_1)+ a_q^2 (\phi, g'_1) \right] \, \delta \left(\hat s - M_{Z'}^2 \right) \,,
\end{equation}
where the $K$-factor represents the enhancement from higher order QCD processes estimated to be $K \simeq 1.3$~\cite{Barger}. After folding $\hat \sigma$ with the CTEQ6 parton distribution functions~\cite{Pumplin:2002vw}, we determine (at the parton level) the resonant production cross section. In Fig.~\ref{LHC-Zprime}  we compare the predicted $\sigma (p\bar p \to Z') \times {\rm BR} (Z' \to \ell \ell)$ (left panel) and $\sigma (p\bar p \to Z') \times {\rm BR} (Z' \to jj)$ (right panel) production rates with  95\% CL upper limits recently reported by the ATLAS~\cite{Collaboration:2011dc} and CMS~\cite{Chatrchyan:2011ns} collaborations. Selection cuts will probably reduced  event rates by factors of 2 to 3. Keeping this in mind, we conclude that the 2012 LHC7 run will probe $3~{\rm TeV} < M_{Z'} < 4~{\rm TeV}$, whereas  future runs from LHC14 will provide a generous discovery potential of up to about $M_{Z'} \sim 8~{\rm TeV}.$

We turn now to discuss the string origin and the compelling properties of the $U(3)_C \times Sp(1)_L \times U(1)_L \times U(1)_R$ gauge group.

\section{Perturbative D-brane Models in a Nutshell} 
\label{S5}

At the time of its formulation and for years thereafter, Superstring
Theory was regarded as a unifying framework for Planck-scale quantum
gravity and TeV-scale SM physics.  Important advances were fueled by
the realization of the vital role played by
D-branes~\cite{Polchinski:1995mt} in connecting string theory to
phenomenology. This has permitted the
formulation~\cite{Antoniadis:1998ig} of string theories with
compositeness setting in at TeV scales and large extra dimensions, see
Appendix~\ref{tev}. There are two paramount phenomenological
consequences for TeV scale D-brane string physics: the emergence of
Regge recurrences at parton collision energies $\sqrt{\hat s} \sim
{\rm string\ scale} \equiv M_s;$ and the presence of one or more
additional $U(1)$ gauge symmetries, beyond the $U(1)_Y$ of the SM.

D-brane TeV-scale string compactifications provide a collection of building block rules that can used to build up the SM or something very close to it~\cite{Blumenhagen:2001te,Kiritsis:2002aj,Kiritsis:2003mc}. The details of the D-brane construct depend a lot on whether we use oriented string or unoriented string models. The basic unit of gauge invariance for oriented string models is a $U(1)$ field, so that a stack of $N$ identical D-branes eventually generates a $U(N)$ theory with the associated $U(N)$ gauge group. In the presence of many D-brane types, the gauge group becomes a product form $\prod U(N_i)$, where $N_i$ reflects the number of D-branes in each stack. Gauge bosons (and associated gauginos in a SUSY model) arise from strings terminating on {\em one} stack of D-branes, whereas chiral matter fields are obtained from strings stretching between {\em two} stacks. Each of the two strings end points carries a fundamental charge with respect to the stack of branes on which it terminates. Matter fields thus posses quantum numbers associated with a bifundamental representation.  In orientifold brane configurations, which are necessary for tadpole cancellation, and thus consistency of the theory, open strings become in general non-oriented. For unoriented strings the above rules still apply, but we are allowed many more choices because the branes come in two different types. There are the branes whose images under the orientifold are different from themselves and their image branes, and also branes who are their own images under the orientifold procedure. Stacks of the first type combine with their mirrors and give rise to $U(N)$ gauge groups, while stacks of the second type give rise to only $SO(N)$ or $Sp(N)$ gauge groups.

The minimal embedding of the SM particle spectrum requires at least
three brane stacks~\cite{Antoniadis:2000ena} leading to three distinct
models of the type $U(3)_C\times U(2)_L\times U(1)$ that were
classified in~\cite{Antoniadis:2000ena, Antoniadis:2004dt}. Only one
of them (model C of~\cite{Antoniadis:2004dt}) has baryon number as
symmetry that guarantees proton stability (in perturbation theory),
and can be used in the framework of TeV strings. Moreover, since $Q_2$
(associated to the $U(1)$ of $U(2)_L$) does not participate in the
hypercharge combination, $U(2)_L$ can be replaced by $Sp(1)_L$ leading
to a model with one extra $U(1)$, the baryon number, besides
hypercharge~\cite{Berenstein:2006pk}. Since baryon number is
anomalous, the extra abelian gauge field becomes massive by the
Green-Schwarz mechanism~\cite{Green:1984sg}, behaving at low energies
as a $Z'$ with a mass in general lower than the string scale by an
order of magnitude corresponding to a loop
factor~\cite{Antoniadis:2002cs}. Lepton number is not a symmetry
creating a problem with large neutrino masses through the Weinberg
dimension-five operator $LLHH$ suppressed only by the TeV string
scale. 

\begin{figure}[tbp] \begin{minipage}[t]{0.49\textwidth} \postscript{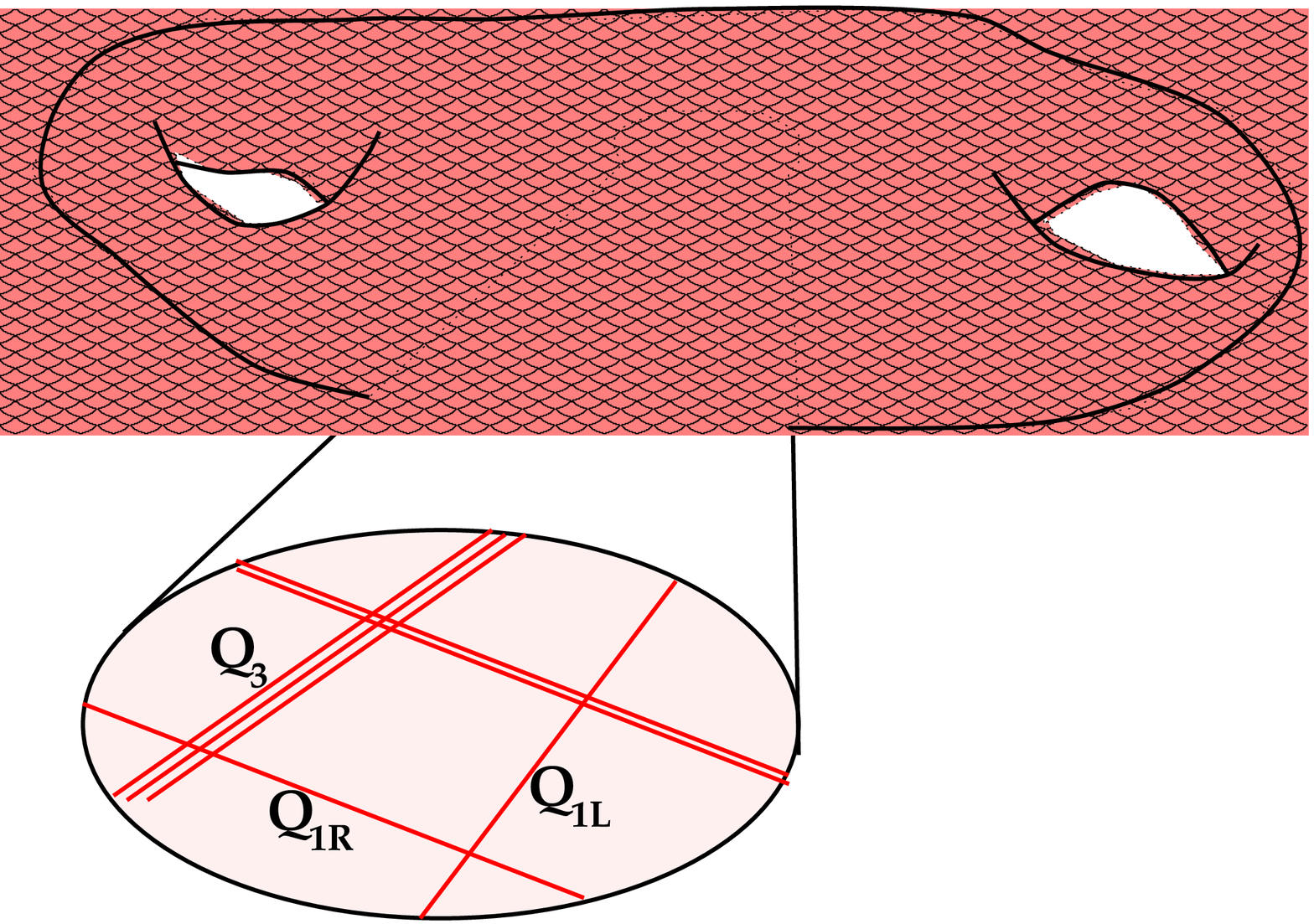}{0.99} \end{minipage} \hfill \begin{minipage}[t]{0.49\textwidth} \postscript{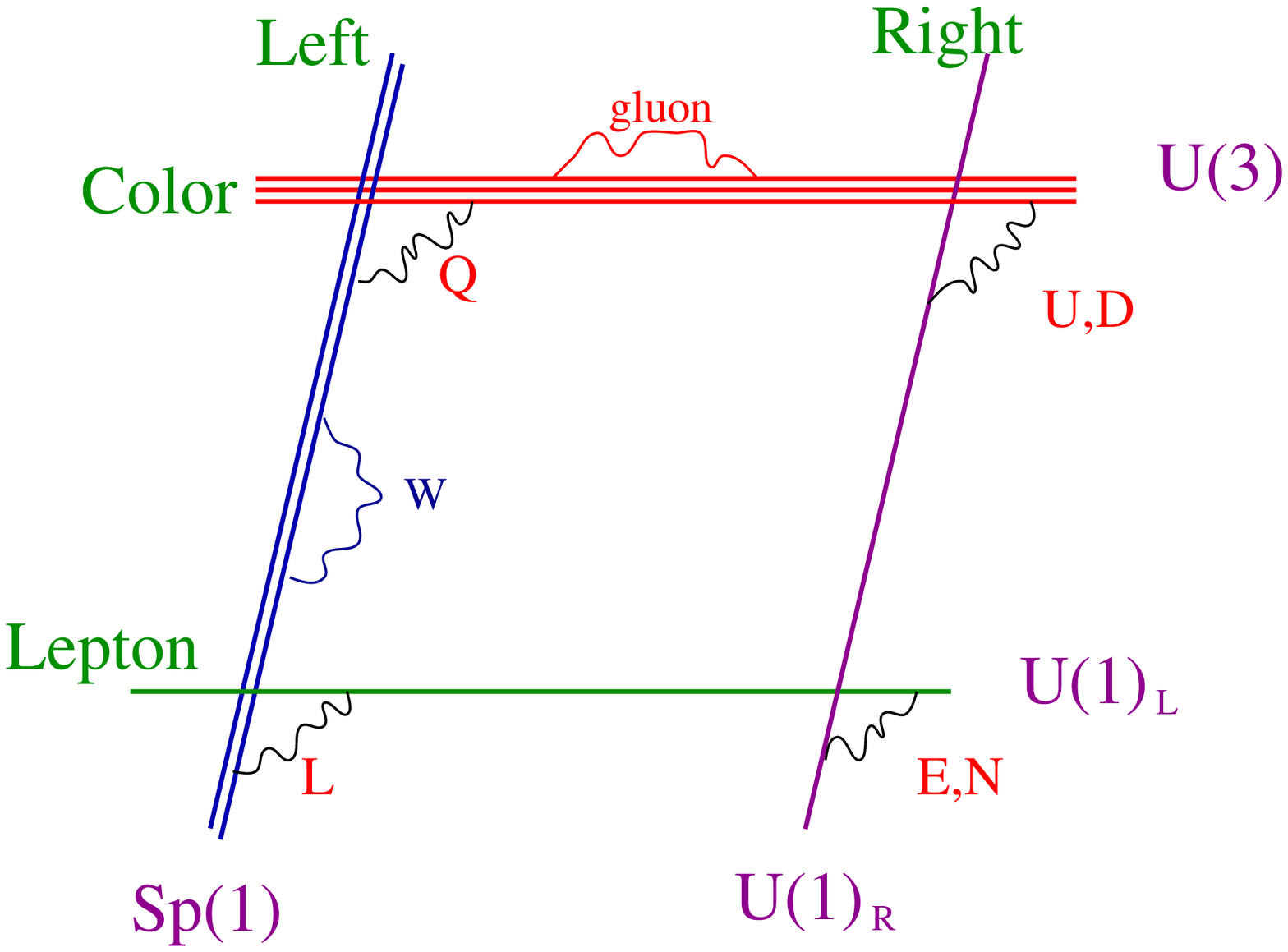}{0.99} \end{minipage} 
\caption{Pictorial representation of the $U(1)_C \times Sp(1)_L \times U(1)_L \times U(1)_R$ D-brane model.}
\label{cartoon}
\end{figure}

The SM embedding in four D-brane stacks leads to many more models that
have been classified in~\cite{Antoniadis:2002qm,
  Anastasopoulos:2006da}. In order to make a phenomenologically
interesting choice, we focus on models where $U(2)_L$ can be reduce to
$Sp(1)$. The minimal SM extension build up out of four stackes of
D-branes is $U(3)_C \times Sp(1)_L \times U(1)_L \times U(1)_R$. A
schematic representation of the D-brane structure is shown in
Fig.~\ref{cartoon}.  The corresponding fermion quantum numbers are
given in Table~\ref{t:spectrum}.  Recall that the combination $B-L$ is
anomaly free. As mentioned already, anomalous $U(1)$'s become massive
necessarily due to the Green-Schwarz anomaly cancellation, but non
anomalous $U(1)$'s can also acquire masses due to effective
six-dimensional anomalies associated for instance to sectors
preserving $N=2$ SUSY~\cite{Antoniadis:2002cs}.\footnote{In
  fact, also the hypercharge gauge boson of $U(1)_Y$ can acquire a
  mass through this mechanism.  In order to keep it massless, certain
  topological constraints on the compact space have to be met.} These
two-dimensional `bulk' masses become therefore larger than the
localized masses associated to four-dimensional anomalies, in the
large volume limit of the two extra dimensions. Specifically for
D$p$-branes with $(p-3)$-longitudinal compact dimensions the masses of
the anomalous and, respectively, the non-anomalous $U(1)$ gauge bosons
have the following generic scale behavior: \ba
{\rm anomalous}~U(1)_i:~~~M_{Z'}&=&g'_iM_s\, ,\nonumber\\
{\rm non-anomalous}~U(1)_i:~~~M_{Z''}&=&g'_iM_s^3\, V_2\, .  \ea Here
$g'_i$ is the gauge coupling constant associated to the group
$U(1)_i$, given by $g'_i\propto g_s/\sqrt{V_\parallel}$ where $g_s$ is
the string coupling and $V_\parallel$ is the internal D-brane
world-volume along the $(p-3)$ compact extra dimensions, up to an
order one proportionality constant. Moreover, $V_2$ is the internal
two-dimensional volume associated to the effective six-dimensional
anomalies giving mass to the non-anomalous $U(1)_i$. E.g. for the case
of $D5$-branes, whose common intersection locus is just 4-dimensional
Minkowski-space, $V_\parallel=V_2$ denotes the volume of the
longitudinal, two-dimensional space along the two internal $D5$-brane
directions.  Since internal volumes are bigger than one in string
units to have effective field theory description, the masses of
non-anomalous $U(1)$-gauge bosons are generically larger than the
masses of the anomalous gauge bosons. Since we want to identify the
light $Z'$ gauge boson with baryon number, which is always anomalous,
a hierarchy compared to the second $U(1)$-gauge boson $Z''$ can arise,
if we identify $Z''$ with the anomaly free combination $B-L$, and take
the internal world-volume $V_2$ a bit larger than the string
scale.\footnote{In \cite{Conlon:2008wa} a different (possibly T-dual)
  scenario with $D7$-branes was investigated. In this case the masses
  of the anomalous and non-anomalous $U(1)$'s appear to exhibit a
  dependence on the entire six-dimensional volume, such that the
  non-anomalous masses become lighter than the anomalous ones.}  In
principle, this hierarchy can be advocated to explain the $Z' - Z''$
mass ratio required to explain the CDF anomaly.\footnote{It is important to stress that in SUSY models derived from D-brane compactifications there can be a light $Z'$ even if the string scale is ${\cal O} (M_{\rm Pl})$~\cite{Mirjam}.}

Particles created by vibrations of relativistic strings populate Regge
trajectories relating their spins $J$ and masses $M$, 
\begin{equation}
J = J_0 + \alpha' M^2 \,,
\end{equation}
where $\alpha' = M_s^{-2}$ is the Regge slope parameter. Thus, if $M_s$ is of order few TeVs, a whole tower of infinite string excitations will open up at this low mass threshold. Should nature be so cooperative, one would expect to see a few string states produced at the LHC. The leading contributions of Regge recurrences to certain processes at hadron colliders are {\em
  universal}. This is because the full-fledged string amplitudes which
describe $2 \to 2$ parton scattering subprocesses involving four gauge
bosons as well as those with two gauge bosons and two chiral matter
fields are (to leading order in string coupling, but all orders in
$\alpha'$) independent of the compactification scheme. Only one
assumption will be necessary in order to set up a solid framework: the
string coupling must be small for the validity of perturbation theory
in the computations of scattering amplitudes. In this case, black hole
production and other strong gravity effects occur at energies above
the string scale (see Appendix~\ref{tev}), therefore at least the few lowest Regge recurrences
are available for examination, free from interference with some
complex quantum gravitational phenomena. We discuss this next.

\section{Regge recurrences} 

The most direct way to compute the amplitude for the scattering of
four gauge bosons is to consider the case of polarized particles
because all non-vanishing contributions can be then generated from a
single, maximally helicity violating (MHV), amplitude -- the so-called
{\it partial\/} MHV amplitude~\cite{Parke:1986gb}.  Assume that two
vector bosons, with the momenta $k_1$ and $k_2$, in the $U(N)$ gauge
group states corresponding to the generators $T^{a_1}$ and $T^{a_2}$
(here in the fundamental representation), carry negative helicities
while the other two, with the momenta $k_3$ and $k_4$ and gauge group
states $T^{a_3}$ and $T^{a_4}$, respectively, carry positive
helicities. (All momenta are incoming.)  Then the partial amplitude
for such an MHV configuration is given
by~\cite{Stieberger:2006te} \begin{equation}
  \label{ampl}
  {\cal A}(A_1^-,A_2^-,A_3^+,A_4^+) ~=~ 4\, g^2\, {\rm Tr}
  \, (\, T^{a_1}T^{a_2}T^{a_3}T^{a_4}) {\langle 12\rangle^4\over
    \langle 12\rangle\langle 23\rangle\langle 34\rangle\langle
    41\rangle}V(k_1,k_2,k_3,k_4)\ ,
\end{equation}
where $g$ is the $U(N)$ coupling constant, $\langle ij\rangle$ are the
standard spinor products written in the notation of
Ref.~\cite{Mangano:1990by}, and the Veneziano formfactor,
\begin{equation}
V(k_1,k_2,k_3,k_4) = V(  s,   t,   u)= \frac{s\,u}{tM_s^2}B(-s/M_s^2,-u/M_s^2)={\Gamma(1-   s/M_s^2)\ \Gamma(1-   u/M_s^2)\over
    \Gamma(1+   t/M_s^2)} \label{formf}
\end{equation}
is the function of Mandelstam variables,
$s=2k_1k_2$, $t=2  k_1k_3$, $u=2 k_1k_4$; $s+t+u=0$.  (For simplicity we drop carets for the parton subprocess.)
The physical content of the form factor becomes clear after using the
well-known expansion in terms of $s$-channel resonances~\cite{Veneziano:1968yb}
\begin{equation}
B(-s/M_s^2,-u/M_s^2)=-\sum_{n=0}^{\infty}\frac{M_s^{2-2n}}{n!}\frac{1}{s-nM_s^2}
\Bigg[\prod_{J=1}^n(u+M^2_sJ)\Bigg],\label{bexp}
\end{equation}
which exhibits $s$-channel poles associated to the propagation of
virtual Regge excitations with masses $\sqrt{n}M_s$. Thus near the
$n$th level pole $(s\to nM^2_s)$:
\begin{equation}\qquad
V(  s,   t,   u)\approx \frac{1}{s-nM^2_s}\times\frac{M_s^{2-2n}}{(n-1)!}\prod_{J=0}^{n-1}(u+M^2_sJ)\ .\label{nthpole}
\end{equation}
In specific amplitudes, the residues combine with the remaining kinematic factors, reflecting the spin content of particles exchanged in the $s$-channel, ranging from $J=0$ to $J=n+1$.\footnote{There are resonances in all the channels, {\em i.e.} there are single particle poles in the $t$ and $u$ channels which would show up as bumps if $t$ or $u$ are positive. However, for physical scattering $t$ and $u$ are negative, so we don't see the bumps.} The low-energy
  expansion reads
\begin{equation}
\label{vexp}
V(s,t,u)\approx 1-{\pi^2\over 6} \frac{s\,
    u}{M_s^4}-\zeta(3)\, \frac{s\, t\, u}{M_s^6}+\dots \, .
\end{equation}

Interestingly, because of the proximity of the 8 gluons and the photon on the color stack of  D-branes, the gluon fusion into $\gamma$ + jet couples at tree level~\cite{Anchordoqui:2007da}.  This implies that there is an order $g^2$ contribution in string theory, whereas this process is not occuring until order $g^4$ (loop level) in field theory. One can write down the total amplitude for this process projecting the gamma ray onto the hypercharge, 
\begin{equation}
  {\cal M} (gg \to \gamma g) = \cos \theta_W \, \, {\cal M} (gg \to Y
  g) = \kappa \, \, \cos \theta_W \, \, {\cal M} (gg \to Cg) \, .
\label{collectableG}
\end{equation}
The $C-Y$ mixing coefficient evaluated at the scale for $U(N)$ unification $M_s$ follows from (\ref{norby}) and is given by
\begin{equation}
\kappa = \frac{c_3 g_Y}{g'_3} = \frac{ g_Y}{\sqrt{6}\, g_3} \, ;  
\end{equation}
it is quite small, around $\kappa \simeq 0.12$ for couplings evaluated
at the $Z$ mass, which is modestly enhanced to $\kappa \simeq 0.14$ as
a result of RG running of the couplings up to $\sim 5$~TeV. 

Consider the amplitude involving three $SU(N)$
gluons $g_1,~g_2,~g_3$ and one $U(1)$ gauge boson $\gamma_4$
associated to the same $U(N)$ stack:
\begin{equation}
\label{gens}
T^{a_1}=T^a \ ,~ \ T^{a_2}=T^b\ ,~ \
  T^{a_3}=T^c \ ,~ \ T^{a_4}=Q \mathbb{I}\ ,
\end{equation}
where $\mathbb{I}$ is the $N{\times}N$ identity matrix and $Q$ is the
$U(1)$ charge of the fundamental representation. 
The color
factor \begin{equation}\label{colf}{\rm
    Tr}(T^{a_1}T^{a_2}T^{a_3}T^{a_4})=Q(d^{abc}+{i\over 4}f^{abc})\ ,
\end{equation}
where the totally symmetric symbol $d^{abc}$ is the symmetrized trace
while $f^{abc}$ is the totally antisymmetric structure constant ~\cite{Mangano:1990by}.

The full MHV amplitude can be obtained~\cite{Stieberger:2006te} by summing
the partial amplitudes (\ref{ampl}) with the indices permuted in the
following way: \begin{equation}
\label{afull} {\cal M}(g^-_1,g^-_2,g^+_3,\gamma^+_4)
  =4\,g^{2}\langle 12\rangle^4 \sum_{\sigma } { {\rm Tr} \, (\,
    T^{a_{1_{\sigma}}}T^{a_{2_{\sigma}}}T^{a_{3_{\sigma}}}T^{a_{4}})\
    V(k_{1_{\sigma}},k_{2_{\sigma}},k_{3_{\sigma}},k_{4})\over\langle
    1_{\sigma}2_{\sigma} \rangle\langle
    2_{\sigma}3_{\sigma}\rangle\langle 3_{\sigma}4\rangle \langle
    41_{\sigma}\rangle }\ ,
\end{equation}
where the sum runs over all 6 permutations $\sigma$ of $\{1,2,3\}$ and
$i_{\sigma}\equiv\sigma(i)$. Note that in the effective field theory
of gauge bosons there are no Yang-Mills interactions that could
generate this scattering process at the tree level. Indeed, $V=1$ at
the leading order of Eq.(\ref{vexp}) and the amplitude vanishes
due to the following identity:
\begin{equation}\label{ymlimit}
{1\over\langle 12\rangle\langle
      23\rangle\langle 34\rangle\langle
      41\rangle}+{1\over\langle 23\rangle\langle
      31\rangle\langle 14\rangle\langle 42\rangle}+{1\over\langle 31\rangle\langle
      12\rangle\langle 24\rangle\langle 43\rangle} ~=~0\ .
\end{equation}
Similarly,
the antisymmetric part of
the color factor (\ref{colf}) cancels out in the full amplitude (\ref{afull}). As a result,
one obtains:
\begin{equation}\label{mhva}
{\cal
    M}(g^-_1,g^-_2,g^+_3,\gamma^+_4)=8\, Q\, d^{abc}g^{2}\langle
  12\rangle^4\left({\mu(s,t,u)\over\langle 12\rangle\langle
      23\rangle\langle 34\rangle\langle
      41\rangle}+{\mu(s,u,t)\over\langle 12\rangle\langle
      24\rangle\langle 13\rangle\langle 34\rangle}\right),
\end{equation}
 where
\begin{equation}
\label{mudef}
\mu(s,t,u)= \Gamma(1-u/M_s^2)\left( {\Gamma(1-s/M_s^2)\over
      \Gamma(1+t/M_s^2)}-{\Gamma(1-t/M_s^2)\over \Gamma(1+s/M_s^2)}\right) .
\end{equation}
All non-vanishing amplitudes can be obtained in a similar way. In particular,
\begin{equation}
\label{mhvb}
{\cal M}(g^-_1,g^+_2,g^-_3,\gamma^+_4)=8\, Q\,
  d^{abc}g^{2}\langle 13\rangle^4\left({\mu(t,s,u)\over\langle
      13\rangle\langle 24\rangle\langle 14\rangle\langle
      23\rangle}+{\mu(t,u,s)\over\langle 13\rangle\langle
      24\rangle\langle 12\rangle\langle 34\rangle}\right),
\end{equation}
and the remaining ones can be obtained either by appropriate
permutations or by complex conjugation.

\begin{figure}
 \postscript{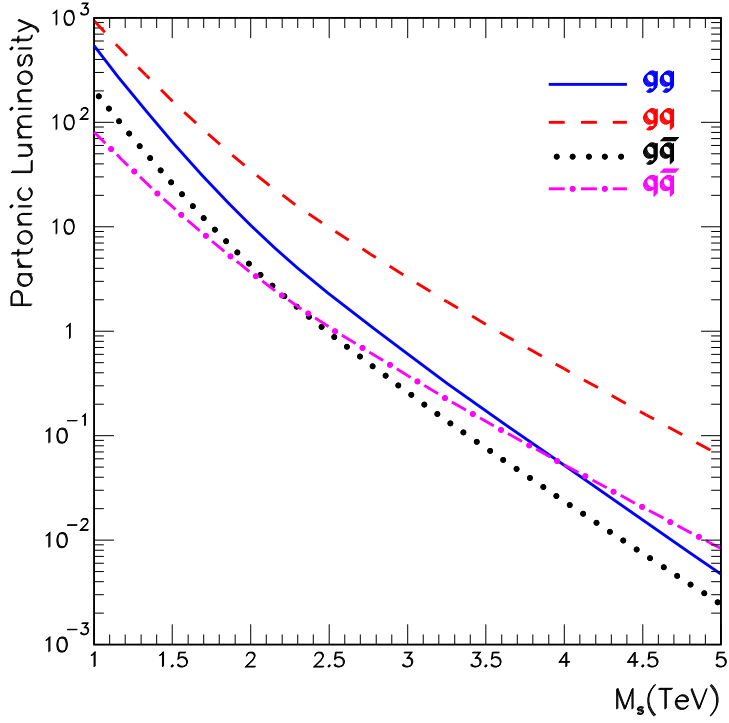}{0.6}
 \caption{Relative contributions of initial state partons ($ij = gg,\ gq,\ g\bar q,\ {\rm and}\ q \bar q$) to $\int_{\tau_0}^1 f_i(x_a, Q) \, \, f_j(\tau_0/x_a, Q)\,\,\, dx_a/x_a$, with varying string scale. From Ref.~\cite{Anchordoqui:2008ac}.}
\label{fig:parton-luminosity}
\end{figure}

In order to obtain the cross section for the (unpolarized) partonic
subprocess $gg\to g\gamma$, we take the squared moduli of individual
amplitudes, sum over final polarizations and colors, and average over
initial polarizations and colors. As an example, the modulus square of
the amplitude (\ref{afull}) is:
\begin{equation}
\label{mhvsq}
|{\cal
    M}(g^-_1,g^-_2,g^+_3,\gamma^+_4)|^2=64\, Q^2\, d^{abc}d^{abc}g^{4}
  \left|{s\mu(s,t,u)\over u}+{s\mu(s,u,t)\over t} \right|^2 \, .
\end{equation}
 Taking
into account all $4(N^2-1)^2$ possible initial polarization/color
configurations and the formula~\cite{vanRitbergen:1998pn}
\begin{equation}
\label{dsq}
\sum_{a,b,c}d^{abc}d^{abc}={(N^2-1)(N^2-4)\over 16 N},
\end{equation}
 we
obtain the average squared amplitude~\cite{Anchordoqui:2007da}
\begin{equation}
\label{mhvav}
|{\cal M}(gg\to
  g\gamma)|^2=g^4Q^2C(N)\left\{ \left|{s\mu(s,t,u)\over
        u}+{s\mu(s,u,t)\over t} \right|^2+(s\leftrightarrow
    t)+(s\leftrightarrow u)\right\},
\end{equation}
 where
\begin{equation}\label{cnn}
C(N)={2(N^2-4)\over N(N^2-1)} \, .
\end{equation}
Before proceeding, we need to make precise the
value of $Q$. If we were considering the process $gg\rightarrow C g,$ then $Q =
\sqrt{1/6}$ due to the normalization condition~(\ref{norm}). However,
for $gg\rightarrow \gamma g$ there are two additional projections given in (\ref{collectableG}):
from $C_\mu$ to the hypercharge boson $Y_\mu$, yielding a mixing factor
$\kappa$; and from $Y_\mu$ onto a photon, providing an additional
factor $\cos\theta_W$. This gives
\begin{equation}
Q = \sqrt{\tfrac{1}{6}} \ \kappa \ \cos \theta_W \, .
\end{equation}

The two most interesting energy regimes of $gg\to g\gamma$
scattering are far below the string mass scale $M_s$
and near the threshold for the production of massive string
excitations. At low energies, Eq.~(\ref{mhvav}) becomes
\begin{equation}
\label{mhvlow}
|{\cal M}(gg\to
  g\gamma)|^2\approx g^4Q^2C(N){\pi^4\over 4 M_s^8}(s^4+t^4+u^4)\qquad
  (s,t,u\ll M_s^2) \, .
\end{equation}
The absence of massless poles, at $s=0$ {\it etc.\/}, translated
into the terms of effective field theory, confirms that there are
no exchanges of massless particles contributing to this process.
On the other hand, near the string threshold $s\approx M_s^2$
\begin{equation}
\label{mhvlow3}
|{\cal M}(gg\to g\gamma)|^2\approx
4g^4Q^2C(N){M_s^8+t^4+u^4\over M_s^4(s-M_s^2)^2}
\qquad (s\approx M_s^2) \,;
\end{equation}
see Appendix~\ref{venpole} for details.

The general form of (\ref{afull})  for any given four external gauge bosons reads
 \begin{eqnarray}
{\cal M}(A_{1}^{-},A_{2}^{-},A_{3}^{+},A_{4}^{+}) & = & 4\, g^{2}\langle12\rangle^{4}\bigg[
\frac{V_{t}}{\langle12\rangle\langle23\rangle\langle34\rangle\langle41\rangle}
\makebox{Tr}(T^{a_{1}}T^{a_{2}}T^{a_{3}}T^{a_{4}}+T^{a_{2}}T^{a_{1}}T^{a_{4}}
T^{a_{3}})\nonumber \\
 & + & \frac{V_{u}}{\langle13\rangle\langle34\rangle\langle42\rangle\langle21\rangle}
 \makebox{Tr}(T^{a_{2}}T^{a_{1}}T^{a_{3}}T^{a_{4}}+T^{a_{1}}T^{a_{2}}T^{a_{4}}
 T^{a_{3}})\nonumber \\
 & + & \frac{V_{s}}{\langle14\rangle\langle42\rangle\langle23\rangle\langle31
 \rangle}\makebox{Tr}(T^{a_{1}}T^{a_{3}}T^{a_{2}}T^{a_{4}}+T^{a_{3}}T^{a_{1}}
 T^{a_{4}}T^{a_{2}})\bigg] , \label{mhv}
\end{eqnarray}
where
\begin{equation}
V_t =V(  s,  t,  u) ~,\qquad V_u=V(  t,  u,  s) ~,\qquad  V_s=V(  u,   s,   t) \, .
\end{equation}
The modulus square of the four-gluon amplitude, summed over final polarizations and colors, and averaged over all $4 (N^2 -1)^2$ possible initial polarization/color configurations follows from (\ref{mhv}) and is given by~\cite{Lust:2008qc} 
\begin{eqnarray}
|{\cal M} (gg \to gg)| ^2 & = & g^4 \left(\frac{1}{  s^2} + \frac{1}{  t^2} + \frac{1}{  u^2} \right) \left[ \frac{2 N^2}{N^2-1} \, (  s^2 \, V_s^2 +   t^2 \, V_t^2 +   u^2 \,  V_u^2) \right. \nonumber \\
 & + & \left. \frac{4 (3 - N^2)}{N^2 (N^2-1)} \, (  s \, V_s +   t \, V_t +   u \, V_u)^2 \right] \, .
\label{gggg}
\end{eqnarray}

The amplitudes involving two gluons and two quarks are also independent of the details of the compactification, such as the configuration of branes, the geometry of the extra dimensions, and whether SUSY is broken or not. This model independence makes it possible to compute all the string corrections to dijet signals at the LHC. The corresponding $2\to 2$ scattering amplitudes, computed at the leading order in
string perturbation theory, are collected in Ref.~\cite{Lust:2008qc}. 
The average square amplitudes are given by the following:
\begin{equation}
|{\cal M} (gg \to q \bar q)|^2 =
g^4 N_f \frac{  t^2 +   u^2}{  s^2} \left[\frac{1}{2N} \frac{1}{  u \,   t} (  t \, V_t +   u \, V_u)^2 - \frac{N}{N^2 -1} \, V_t \, V_u \right] \,\,,
\label{ggqbarq}
\end{equation}
\begin{equation}
|{\cal M} (q \bar q \to gg)|^2 =
g^4 \ \frac{  t^2 +   u^2}{  s^2} \ \left[\frac{(N^2 -1)^2}{2 N^3} \ \frac{1}{  u   t} \ (  t \, V_t +   u \, V_u)^2 - \frac{N^2 -1}{N} \, V_t \, V_u \right] \, \,,
\label{qbarqgg}
\end{equation}
and 
\begin{equation}
|{\cal M}(qg \to qg)|^2 =
g^4 \ \frac{  s^2 +   u^2}{  t^2} \left[V_s \, V_u - \frac{N^2 -1}{2 N^2} \ \ \frac{1}{  s   u} \ (  s\, V_s +   u \, V_u)^2 \right] \,\, .
\label{qgqg}
\end{equation}

The amplitudes for the four-fermion processes like quark-antiquark
scattering are more complicated because the respective form factors
describe not only the exchanges of Regge states but also of heavy
Kaluza-Klein and winding states with a model-dependent spectrum
determined by the geometry of extra dimensions. Fortunately, they are
suppressed, for two reasons. First, the QCD $SU(3)$ color group
factors favor gluons over quarks in the initial state. Second, the
parton luminosities in proton-proton collisions at the LHC, at the
parton center of mass energies above~1 TeV, are significantly lower
for quark-antiquark subprocesses than for gluon-gluon and
gluon-quark, see Fig.~\ref{fig:parton-luminosity}. The collisions of valence
quarks occur at higher luminosity; however, there are no Regge
recurrences appearing in the $s$-channel of quark-quark
scattering~\cite{Lust:2008qc}.

In the following we isolate the contribution from the first resonant
state in Eqs.~(\ref{gggg}) - (\ref{qgqg}). For partonic center of mass
energies $\sqrt{s}<M_s$, contributions from the Veneziano functions
are strongly suppressed, as $\sim (\sqrt{s}/M_s)^8$, over standard
model processes; see Eq.~(\ref{mhvlow}). (Corrections to SM processes
at $\sqrt{s} \ll M_s$ are of order $(\sqrt{s}/M_s)^4$; see
Eq.~(\ref{vexp}).) In order to factorize amplitudes on the poles due
to the lowest massive string states, it is sufficient to consider
$s=M_s^2$. In this limit, $V_s$ is regular while
\begin{equation}
V_t \to \frac{  u}{  s-M_s^2}~,\qquad V_u \to \frac{  t}{  s-M_s^2} \, .
\end{equation}
Thus the $s$-channel pole term of the average square amplitude
(\ref{gggg}) can be rewritten as
\begin{equation}
|{\cal M} (gg \to gg)| ^2  =  2 \ 
\frac{g^4}{M_s^4}\ \left(\frac{N^2-4+(12/N^2)}{N^2-1}\right) 
 \ \frac{M_s^8+  t^4 +   u^4}{(  s - M_s^2)^2} \, .
\label{ggggpole}
\end{equation}
Note that the contributions of single poles to the cross section are
antisymmetric about the position of the resonance, and vanish in any
integration over the resonance.\footnote{As an illustration, consider the
  amplitude $a +b/D$ in the vicinity of the pole, where $a$ and $b$
  are real, $D = x+i\epsilon,$ $x=s-M_s^2,$ and $\epsilon = \Gamma
  M_s.$ Then, since Re$(1/D) = x/|D|^2$, the cross section becomes
  $\sigma \propto a^2 + b^2/|D|^2 + 2 \, a \, b \, x/|D|^2 \simeq a^2
  + b^2 \, \pi \, \delta(x)/\epsilon + 2ab \, \pi \, x \
  \delta(x)/\epsilon$. Integrating over the width of the resonance,
  one obtains $a^2 \epsilon + b^2 \pi/\epsilon \simeq b \pi$, because
  $b \propto \epsilon$, $a \propto g^2$ and $\epsilon \propto g^2$.}
Before proceeding, we pause to present our notation. The first Regge
excitations of the gluon $g$, the color singlet $C$, and quarks $q$
will be denoted by $g^*,\ C^*,\, q^*$, respectively. Recall that
$C_\mu$ has an anomalous mass in general lower than the string scale
by an order of magnitude. If that is the case, and if the mass of the
$C^*$ is composed (approximately) of the anomalous mass of the $C_\mu$
and $M_s$ added in quadrature, we would expect only a minor error in
our results by taking the $C^*$ to be degenerate with the other
resonances.  The singularity at $s=M_s^2$ needs softening to a
Breit-Wigner form, reflecting the finite decay widths of resonances
propagating in the $s$ channel. Due to averaging over initial
polarizations, Eq.(\ref{ggggpole}) contains additive contributions
from both spin $J=0$ and spin $J=2$ $U(3)$ bosonic Regge excitations
($g^*$ and $C^*$), created by the incident gluons in the helicity
configurations ($\pm \pm$) and ($\pm \mp$), respectively.  The $M_s^8$
term in Eq.~(\ref{ggggpole}) originates from $J=0$, and the $ t^4+
u^4$ piece reflects $J=2$ activity. Since the resonance widths depend
on the spin and on the identity of the intermediate state ($g^*$,
$C^*$) the pole term (\ref{ggggpole}) should be smeared
as~\cite{Anchordoqui:2008di}
\begin{eqnarray}
\label{gggg2}
|{\cal M} (gg \to gg)| ^2 & = & 2\ 
\frac{g^4}{M_s^4}\ \left(\frac{N^2-4+(12/N^2)}{N^2-1}\right)  \\
 & \times & \left\{ W_{g^*}^{gg \to gg} \, \left[\frac{M_s^8}{(  s-M_s^2)^2 
+ (\Gamma_{g^*}^{J=0}\ M_s)^2} \right. \right.
\left. +\frac{  t^4+   u^4}{(  s-M_s^2)^2 + (\Gamma_{g^*}^{J=2}\ M_s)^2}\right] \nonumber \\
   & + &
W_{C^{O*}}^{gg \to gg} \, \left. \left[\frac{M_s^8}{(  s-M_s^2)^2 + (\Gamma_{C^*}^{J=0}\ M_s)^2} \right.
\left. +\frac{  t^4+  u^4}{(  s-M_s^2)^2 + (\Gamma_{C^*}^{J=2}\ M_s)^2}\right] \right\}, \nonumber
\end{eqnarray}
where $\Gamma_{g^*}^{J=0} = 75\, (M_s/{\rm TeV})~{\rm GeV}$,
$\Gamma_{C^{*}}^{J=0} = 150 \, (M_s/{\rm TeV})~{\rm GeV}$,
$\Gamma_{g^*}^{J=2} = 45 \, (M_s/{\rm TeV})~{\rm GeV}$, and
$\Gamma_{C^{*}}^{J=2} = 75 \, (M_s/{\rm TeV})~{\rm GeV}$ are the
total decay widths for intermediate states $g^*$ and $C^*$, with
angular momentum $J$~\cite{Anchordoqui:2008hi}.  The associated weights of these intermediate states are given in terms of the probabilities for the various entrance and exit channels
\begin{eqnarray} 
\el{totalcrossdecom} \frac{N^2-4+12/N^2}{N^2-1} & = & \frac {16} {(N^2-1)^2}\left[\left(N^2-1\right)\left(\frac{N^2-4}{ 4N}\right)^2+
  \left(\frac{N^2-1}{2N}\right)^2\right] \nonumber \\ & \propto & \frac {16} {(N^2-1)^2}
\left[(N^2-1)(\Gamma_{g^*\to gg})^2 + (\Gamma_{C^*\to gg})^2\right] \,,
\end{eqnarray}
yielding
\begin{equation}
W_{g^*}^{gg \to gg} = \frac{8 (\Gamma_{g^* \to gg})^2}{8(\Gamma_{g^* \to gg})^2 +
(\Gamma_{C^* \to gg})^2} = 0.44, 
\label{w1}
\end{equation}
and
\begin{equation}
W_{C^*}^{gg \to gg} = \frac{(\Gamma_{C^*
  \to gg})^2}{8(\Gamma_{g^* \to gg})^2 + (\Gamma_{C^* \to gg})^2} =0.56  \, .
\label{w2}
\end{equation}
A similar calculation transforms Eq.~(\ref{ggqbarq}) near the pole into
\begin{eqnarray}
  |{\cal M} (gg \to q \bar q)|^2 & = & \frac{g^4}{M_s^4}\ N_f\ \left(\frac{N^2-2}{N(N^2-1)}\right)
  \left [W_{g^*}^{gg \to q \bar q}\, \frac{  u   t(   u^2+   t^2)}{(  s-M_s^2)^2 + (\Gamma_{g^*}^{J=2}\ M_s)^2} \right. \nonumber \\
  & + & \left. W_{C^{*}}^{gg \to q \bar q}\, \frac{  u   t (   u^2+   t^2)}{(  s-M_s^2)^2 + 
      (\Gamma_{C^{*}}^{J=2}\ M_s)^2} \right] \, ,
\end{eqnarray}
where 
\begin{equation}
W_{g^*}^{gg \to q \bar q}  = W_{g^*}^{q \bar q \to gg} = 
\frac{8\Gamma_{g^* \to gg} \, 
\Gamma_{g^* \to q \bar q}} {8\Gamma_{g^* \to gg} \, 
\Gamma_{g^* \to q \bar q} + \Gamma_{C^{*} \to gg} \, 
\Gamma_{C^{*} \to q \bar q}}  = 0.71
\label{w3}
\end{equation}
 and 
\begin{equation}
W_{C^{*}}^{gg \to q \bar q} = W_{C^{*}}^{q \bar q \to gg}  = 
\frac{\Gamma_{C^{*} \to gg} \, 
\Gamma_{C^{*} \to q \bar q}}{8\Gamma_{g^* \to gg} \, 
\Gamma_{g^* \to q \bar q} + \Gamma_{C^{*} \to gg} \, 
\Gamma_{C^{*} \to q \bar q}}  = 0.29\, .
\label{w4}
\end{equation}
Near the $  s$ pole Eq.~(\ref{qbarqgg}) becomes 
\begin{eqnarray}
|{\cal M} (q \bar q \to gg)|^2  & = &  \frac{g^4}{M_s^4}\ \left(\frac{(N^2 -2) 
(N^2-1)}{N^3}\right)
\left[ W_{g^*}^{q\bar q \to gg} \,  \frac{  u   t(   u^2+   t^2)}{(  s-M_s^2)^2 + (\Gamma_{g^*}^{J=2}\ M_s)^2} \right. \nonumber \\
 & + & \left.  W_{C^{*}}^{q\bar q \to gg} \, \frac{  u   t(   u^2+   t^2)}{(  s-M_s^2)^2 + (\Gamma_{C^{*}}^{J=2}\ M_s)^2} \right] \,\,,
\end{eqnarray}
whereas Eq.~(\ref{qgqg}) can be rewritten as
\begin{eqnarray}
|{\cal M}(qg \to qg)|^2 & = & - \frac{g^4}{M_s^2}\ 
\left(\frac{N^2-1}{2N^2}\right)
\left[ \frac{M_s^4   u}{(  s-M_s^2)^2 + (\Gamma_{q^*}^{J=1/2}\ M_s)^2}\right. \nonumber \\
 & + & \left. \frac{  u^3}{(s-M_s^2)^2 + (\Gamma_{q^*}^{J=3/2}\ M_s)^2}\right] \, \, .
\label{qgqg2}
\end{eqnarray}
The total decay widths for the $q^*$ excitation are: $\Gamma_{q^*}^{J=1/2} = \Gamma_{q^*}^{J=3/2} = 37\, (M_s/{\rm TeV})~{\rm GeV}$~\cite{Anchordoqui:2008hi}.  Superscripts $J=2$ are understood to be inserted on all the $\Gamma$'s in Eqs.~(\ref{w1}), (\ref{w2}), (\ref{w3}), (\ref{w4}).  Equation~(\ref{gggg2}) reflects the fact that weights for $J=0$ and $J=2$ are the same~\cite{Anchordoqui:2008hi}.

The $s$-channel poles near the second Regge resonance can be
approximated by expanding the Veneziano formfactor $V_t$ around $s = 2
M_s^2$,
\begin{equation}
V(s,t,u) \approx \frac{u (u +M_s^2)}{M_s^2 (s - 2 M_s^2)} \, .
\end{equation}
The associated scattering amplitudes and decay widths of the $n=2$
string resonaces are collected in Ref.~\cite{Dong:2010jt}. Generally,
the width of the Regge excitations will grow at least linearly with
energy, whereas the spacing between levels will decrease with
energy. This implies an upper limit on the domain of validity for this
phenomenological approach~\cite{Cornet:2001gy}. In particular, for a resonance $R$ of mass
$M$, the total width is given by
\begin{equation}
\Gamma_{\rm tot} \sim \frac{g^2}{4 \, \pi} \, {\cal C} \, \frac{M}{4},
\end{equation}
 where ${\cal C} > 1$ because of the growing multiplicity of decay modes~\cite{Anchordoqui:2008hi,Dong:2010jt}. On the other hand, since $\Delta (M^2) = M_s^2$ the level spacing at mass $M$ is $\Delta M \sim M_s^2/(2M)$; thus,
\begin{equation}
\frac{\Gamma_{\rm tot}}{\Delta M} \sim \frac{g^2}{8 \pi} \ {\cal C} \ \left(\frac{M}{M_s} \right)^2 = \frac{g^2}{8 \pi} \ {\cal C} \ n <1 \, .
\end{equation}
For excitation of the resonance $R$ via $a+b\rightarrow R$, the assumption $\Gamma_{\rm tot}(R) \sim \Gamma(R\rightarrow ab)$ (which underestimates the real width) yields a perturbative regime for $n \alt 40$. This is to be compared with the $n \sim 10^4$ levels of the string needed for black hole production (see Appendix~\ref{tev}).

Given the particular nature of the process we are considering, the production of a TeV-scale resonance and its subsequent 2-body decay, several observables at the LHC are available. Most apparently, one would hope that the resonance would  be visible in data binned
according to the dijet invariant mass $M$, after setting cuts on the
different jet rapidities, $|y_1|, \, |y_2| \le y_{\rm max}$ and transverse
momenta $p_{\rm T}^{1,2}>50$~GeV.  With the definitions $Y\equiv
\frac{1}{2} (y_1 + y_2)$ and $y \equiv \frac{1}{2} (y_1-y_2)$, the
cross section per interval of $M$ for $p p\rightarrow {\rm dijet}$ is
given by~\cite{Eichten:1984eu}
\begin{eqnarray}
\frac{d\sigma}{dM} & = & M\tau\ \sum_{ijkl}\left[
\int_{-Y_{\rm max}}^{0} dY \ f_i (x_a,\, M)  \right. \ f_j (x_b, \,M ) \
\int_{-(y_{\rm max} + Y)}^{y_{\rm max} + Y} dy
\left. \frac{d\sigma}{d\hat t}\right|_{ij\rightarrow kl}\ \frac{1}{\cosh^2
y} \nonumber \\
& + &\int_{0}^{Y_{\rm max}} dY \ f_i (x_a, \, M) \
f_j (x_b, M) \ \int_{-(y_{\rm max} - Y)}^{y_{\rm max} - Y} dy
\left. \left. \frac{d\sigma}{d\hat t}\right|_{ij\rightarrow kl}\
\frac{1}{\cosh^2 y} \right] \,,
\label{longBH}
\end{eqnarray}
where $f(x,M)$'s are parton distribution functions (we use  CTEQ6~\cite{Pumplin:2002vw}),
$\tau = M^2/s$, $x_a = \sqrt{\tau} e^{Y}$,  $x_b = \sqrt{\tau} e^{-Y},$ and
\begin{equation}
  |{\cal M}(ij \to kl) |^2 = 16 \pi \hat s^2 \,
  \left. \frac{d\sigma}{d\hat t} \right|_{ij \to kl} \, ;
\end{equation}
we reinstate the caret notation $(\hat s, \, \hat t, \, \hat u)$ to specify partonic subprocesses.
The $Y$ integration range in Eq.~(\ref{longBH}), $Y_{\rm max} = {\rm min} \{ \ln(1/\sqrt{\tau}),\ \ y_{\rm max}\}$, comes from requiring $x_a, \, x_b < 1$ together with the rapidity cuts $y_{\rm min} <|y_1|, \, |y_2| < y_{\rm max}$. The kinematics of the scattering also provides the relation $M = 2p_T \cosh y$, which when combined with $p_T = M/2 \ \sin \theta^* = M/2 \sqrt{1-\cos^2 \theta^*},$ yields $\cosh y = (1 - \cos^2 \theta^*)^{-1/2},$ where $\theta^*$ is the center-of-mass scattering angle.  Finally, the Mandelstam invariants occurring in the cross section are given by $\hat s = M^2,$ $\hat t = -\frac{1}{2} M^2\ e^{-y}/ \cosh y,$ and $\hat u = -\frac{1}{2} M^2\ e^{+y}/ \cosh y.$
In what follows we set $N=3$ and $N_f =6$.

The CMS Collaboration has searched for such narrow resonances in their dijet mass spectrum using data from $pp$ collisions at $\sqrt{s} = 7~{\rm TeV}$~\cite{Khachatryan:2010jd}. After operating for only few months, with merely 2.9 inverse picobarns of integrated luminosity, the LHC CMS experiment has ruled out $M_s < 2.5~{\rm TeV}$. The LHC7 has recently delivered an integrated luminosity in excess of $1~{\rm fb}^{-1}$. This extends considerably the search territory for new physics in events containing dijets. The new data exclude string resonances with $M_s < 4~{\rm TeV}$~\cite{Chatrchyan:2011ns}. In fact, the LHC has the capacity of discovering strongly interacting resonances via dijet final states in practically all range up to $\frac{1}{2}\sqrt{s}_{\rm LHC}$. We discuss this next.

\begin{figure}[tpb]
 \postscript{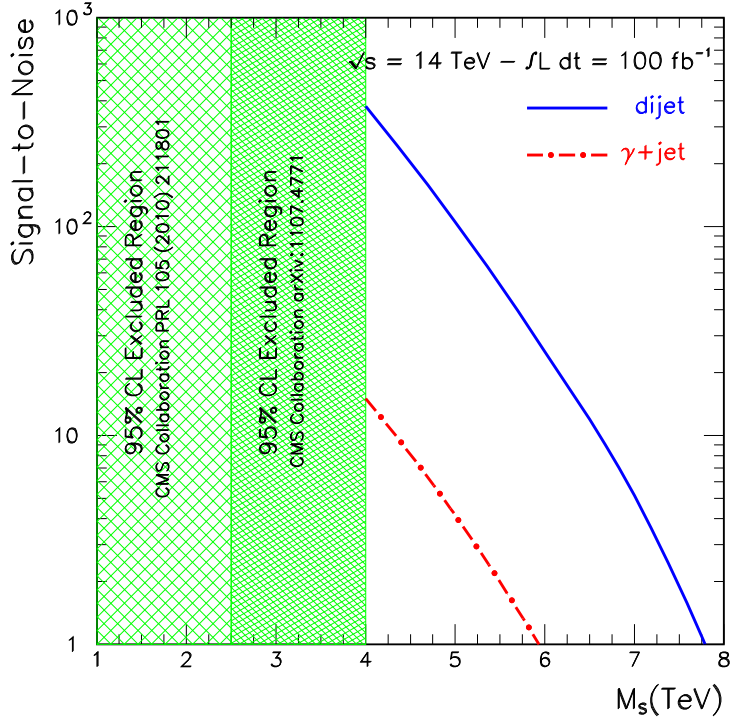}{0.6}
\caption{Signal-to-noise ratio of $pp \to {\rm
    dijet}$ and $pp \to \gamma + {\rm jet}$, for $\sqrt{s} = 14$~TeV, ${\cal L} = 100~{\rm fb}^{-1}$, and  $\kappa^2
  \simeq 0.02$. The approximate
  equality of the background due to misidentified $\pi^0$'s and the
  QCD background, across a range of large $p_T^\gamma$,  is maintained as an approximate
  equality over a range of  $\gamma$-jet invariant masses
  with the rapidity cuts imposed ($|y_{\rm max}^{j}| < 1.0$ and $|y_{\rm max}^{\gamma}| < 2.4$). The 95\%~CL lower limits on the string scale recently reported by the CMS Collaboration are also shown.}
\label{fig:S2N}
\end{figure}

Standard bump-hunting methods, such as calculating cumulative cross sections
\begin{equation}
\sigma (M_0) = \int_{M_0}^\infty  \frac{d\sigma}{dM} \, \, dM
\end{equation}
and searching for regions with significant deviations from the QCD
background, may allow to find an interval of $M$ suspected of
containing a bump.  With the establishment of such a region, one may
calculate a signal-to-noise ratio, with the signal rate estimated in
the invariant mass window $[M_s - 2 \Gamma, \, M_s + 2 \Gamma]$.  To
accommodate the minimal acceptance cuts on dijets from LHC
experiments~\cite{Bhatti:2008hz}, an additional kinematic cut,
$|y_{\rm max}|<1.0$, has been included in the calculation. The noise
is defined as the square root of the number of QCD background events
in the same dijet mass interval for the same integrated luminosity.
Our significant results are encapsuled in Fig.~\ref{fig:S2N}, where we
show the signal-to-noise ratio as a function of the string scale. It
is remarkable that within one to two years of data collection with LHC14, {\em
  string scales as large as $6.8~{\rm TeV}$ are  open to discovery at the $\geq 5\sigma$ level.} Once more, we stress that these results
contain no unknown parameters. They depend only on the D-brane
construct for the SM, and {\it are independent of
  compactification details.}\footnote{The only remnant
of the compactification is the relation between the Yang-Mills
coupling and the string coupling. We take this relation to reduce to
field theoretical results in the case where they exist, {\em e.g.} $gg \to
gg$. Then, because of the require correspondence with field theory,
the phenomenological results are independent of the compactification
of the transverse space. However, a different phenomenology would
result as a consequence of warping one~\cite{Randall:1999ee}  or more parallel
dimensions~\cite{Hassanain:2009at}.}

Although the expected discovery reach would not be  as high as that for dijets, the measurement of $pp\rightarrow\gamma~ +$ jet can potentially provide an interesting corroboration for the stringy origin for new physics manifest as a resonant structure in LHC data. The  Breit-Wigner form  for gluon fusion into $\gamma$ + jet follows from (\ref{mhvlow3}) and is given by
\begin{equation}
\label{mhvlow2}
|{\cal M}(gg\to g\gamma)|^2\simeq
\frac{5g_3^4Q^2}{3M_s^4}\bigg[{M_s^8\over (\hat s-M_s^2)^2+(\Gamma_{g^*}^{J=0} M_s)^2}+
{\hat t^4+\hat u^4\over (\hat s-M_s^2)^2+(\Gamma_{g^*}^{J=2} M_s)^2}\bigg].
\end{equation}
From Fig.~\ref{fig:parton-luminosity} we see that the dominant
$s$-channel pole term of the average square amplitude contributing to
$pp \to \gamma$ + jet is~\cite{Anchordoqui:2009ja}
\begin{equation}
|{\cal M}(qg \to q \gamma)|^2   =   -\frac{g_3^4 Q^2}{3M_s^2}\
\left[ \frac{M_s^4  \hat  u}{( \hat s-M_s^2)^2 + (\Gamma_{q^*}^{J=\frac{1}{2}}\ M_s)^2}  \right.  + \left. \frac{\hat u^3}{(\hat s-M_s^2)^2 + (\Gamma_{q^*}^{J=\frac{3}{2}}\ M_s)^2}\right] \, .
\label{qgqz}
\end{equation}
We duplicate the calculation of the signal-to-noise ratio substituting
in (\ref{longBH}) $d\sigma/d\hat t|_{ij \to kl}$ by $d\sigma/d\hat
t|_{ij \to k\gamma}$. For photons, we set $y_{\rm max} <
2.4$~\cite{Ball:2007zza}. To minimize misidentification with a
high-$p_T \ \pi^0$, isolation cuts must be imposed on the photon, and
to trigger on the desired channel, the hadronic jet must be
identified. A detailed study of the CMS potential for isolation of
prompt-$\gamma$'s has been carried out~\cite{Gupta:2007cy}, using
GEANT4 simulations of $\gamma + {\rm jet}$ events generated with
Pythia. This analysis (which also includes $\gamma$'s produced in the
decays of $\eta$, $K_s^0$, $\omega^0,$ and bremsstrahlung photons
emerging from high-$p_T$ jets) suggests
\begin{equation}
 \beta = \frac{{\rm background \, due \, to \, misidentified} \, \pi^0 \, {\rm
after \, isolation \, cuts}}{{\rm QCD \, background \, from \, direct \, photon \, production}}  + 1  \simeq 2\,\, .
\end{equation}
Therefore, in our numerical calculation we assume the noise is increased by a factor of $\sqrt{\beta}$, over the direct photon QCD contribution.  The signal used to obtain the results displayed in Fig.~\ref{fig:S2N} includes the parton subprocesses $gg \to g \gamma$ (which does not exist at tree level in QCD), $qg \to q \gamma$, $\bar q g \to \bar q \gamma$, and $q \bar q \to g \gamma$. All except the first have been calculated in QCD and constitute the SM background.  {\em For string scales as high as 5.0~{\rm TeV}, observations of resonant structures in $pp\to \gamma + {\rm jet}$ can provide interesting corroboration for stringy physics at the TeV-scale.}

Events with a single jet plus missing energy ($\met$) with balancing transverse momenta (so-called ``monojets'') are incisive probes of new physics. As in the SM, the source of this topology is $ij \to k Z$ followed by $Z \to \nu \bar \nu.$ Both in the SM and string theory the cross section for this process is of order $g^4$. Virtual KK graviton emission ($ij \to k G$) involves emission of closed strings, resulting in an additional suppression of order $g^2$ compared to $Z$ emission. A careful discussion of this suppression is given in~\cite{Cullen:2000ef}. Ignoring the $Z$-mass ({\em i.e.} keeping only transverse $Z$'s), the Regge contribution to $pp \to Z + {\rm jet}$ is suppressed relative to the $pp \to \gamma+ {\rm jet}$ by a factor of $\tan^2\theta_W = 0.29$~\cite{Anchordoqui:2009mm}.

\begin{figure}[tbp]
\postscript{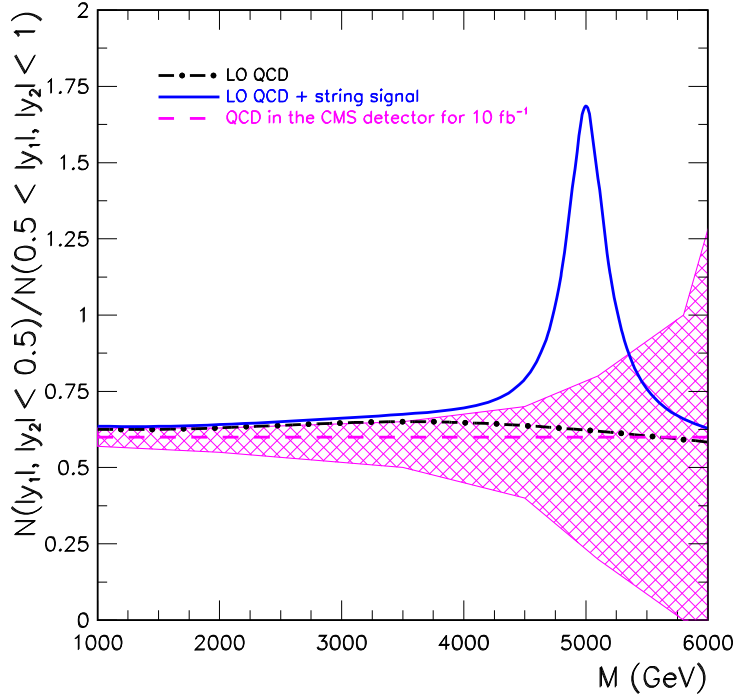}{0.6}
\caption{For a luminosity of 10~fb$^{-1}$, the expected value (dashed
  line) and statistical error (shaded region) of the dijet ratio of
  QCD in the CMS detector is compared with LO QCD (dot-dashed line)
  and LO QCD plus lowest massive string excitation at a scale $M_s =
  5~{\rm TeV}$. From Ref.~\cite{Anchordoqui:2008di}.}
\label{R_CMS10}
\end{figure}

We now turn to the analysis of the dijet angular distributions. QCD
parton-parton cross sections are dominated by $t$-channel exchanges 
that produce dijet angular distributions which peak at small center of
mass scattering angles. In contrast, non--standard contact
interactions or excitations of resonances result in a more isotropic
distribution. In terms of rapidity variables for standard transverse
momentum cuts, dijets resulting from QCD processes will preferentially
populate the large rapidity region, while the new processes generate
events more uniformly distributed in the entire rapidity region. To
analyze the details of the rapidity space the D\O\
Collaboration~\cite{Abbott:1998wh} introduced a new parameter,
\begin{equation}
R = \frac{d\sigma/dM|_ {(|y_1|,|y_2|< 0.5)}}{d\sigma/dM|_{(0.5 < |y_1|,|y_2| < 1.0)}} \, ,
\end{equation}
the ratio of the number of events, in a given dijet mass bin, for both
rapidities $|y_1|, |y_2| < 0.5$ and both rapidities $0.5 < |y_1|,
|y_2| < 1.0$.  In
Fig.~\ref{R_CMS10} we compare the results from a full CMS detector
simulation of the ratio $R$, with predictions from LO QCD
and model-independent contributions to the $q^*$, $g^*$ and $C^*$
excitations.\footnote{An illustration of the use of this parameter in
  a heuristic model where SM amplitudes are modified by a
  Veneziano formfactor has been presented ~\cite{Meade:2007sz}.}   The synthetic
population was generated with Pythia, passed through the full CMS
detector simulation and reconstructed with the ORCA reconstruction
package~\cite{Esen}. For an integrated luminosity
of 10~fb$^{-1}$ the LO QCD contributions with $\alpha_s= g^2_3/4\pi = 0.1$
(corresponding to running scale $\Lambda \approx M_s$) are within
statistical fluctuations of the full CMS detector simulation. (Note
that the string scale is an optimal choice of the running scale which
should normally minimize the role of higher loop corrections.) Since
one of the purposes of utilizing NLO calculations is to fix the choice
of the running coupling, we take this agreement as rationale to omit
loops in QCD and in string theory. It is clear from Fig.~\ref{R_CMS10}
that incorporating NLO calculation of the background and the signal
would not significantly change the large deviation of the string
contribution from the QCD background. String scales $\sim 5~{\rm TeV}$ can be probed with $10~{\rm fb}^{-1}$ of LHC14 data collection. Because of background reduction by optimized  rapidity cuts, the $R$ parameter can (in principle) extend the LHC discovery reach of Regge excitations.

In closing, we note that for $e^+ e^-$ colliders string theory predicts the {\em precise} value, equal to  1/3, of the relative weight of spin 2 and spin 1 contributions~\cite{Anchordoqui:2010zs}. This yields a dimuon angular distribution with a pronounced forward-backward asymmetry, which could help distinguishing between low mass strings and other beyond SM scenarios.

\section{Conclusions}

We have considered a low-mass string compactification in which the SM gauge multiplets originate in open strings ending on D-branes, with gauge bosons due to strings attached to stacks of D-branes and chiral matter due to strings stretching between intersecting D-branes. For the non-abelian $SU(3)$ group, the D-brane construct requires the existence of an additional $U(1)$ gauge boson coupled to baryon number. In this framework, $U(1)$ and $SU(3)$ appear as subgroups of $U(3)$. In addition, our minimal model contains three other stacks of D-branes to accommodate the electroweak $Sp(1)$ left and $U(1)$ fields attached to the lepton D-brane and to the right D-brane. One linear combination of the three $U(1)$ gauge bosons is identified as the the hypercharge $Y$ field, coupled to the anomaly free hypercharge current. The two remaining linear combinations ($Y', Y''$) of the three $U(1)$'s, which can be naturally associated with $B$ and $B-L$, grow masses. After electroweak breaking, mixing with the third component of isospin results in the three observable gauge bosons, where with small mixing $Z'\simeq Y', \, Z''\simeq Y''$.

In our phenomenological discussion about the possible discovery of
massive $Z'$-gauge bosons, we have taken as a benchmark scenario the dijet
plus $W$ signal, recently observed by the CDF Collaboration at
Tevatron. For a fixed $M_{Z'} \simeq 150~{\rm GeV}$, the model is
quite constrained. Fine tuning  its free parameters is just
sufficient to simultaneously ensure: a small $Z-Z'$ mixing in accord
with the stringent LEP data on the $Z$ mass; very small (less than
1\%) branching ratio into leptons; and a large hierarchy between $Z''$
and $Z'$ masses.    

If the CDF anomaly does not survive additional scrutiny (as indicated by the more recent  D\O\ results), the analysis presented here can be directly applied to the higher energy realm, with a view toward identifying the precise makeup of the various abelian sectors, and pursuing with strong confidence a signal at LHC for Regge excitations of the string. 

In D-brane constructions, the full-fledged string amplitudes supplying the dominant contributions to $pp$ scattering cross sections are completely independent of the details of compactification. We have made use of the amplitudes evaluated near the first resonant pole to report on the discovery potential at the LHC for the first Regge excitations of the quark and gluon. The precise predictions for the branching fraction of two different topologies (dijet and $\gamma +$ jet) can be used as a powerful discriminator of low mass string excitations from other beyond SM scenarios. We have long imagined strings to be minuscule objects which could only be experimentally observed in the far-distant future. It is conceivable that this future has already arrived.

\section*{Acknowledgments}

I'm thankful to Ignatios Antoniadis, Vic Feng, Haim Goldberg, Xing Huang,
Dieter L\"ust, Satoshi Nawata, Stephan Stieberger, and Tom Taylor for 
fruitful and enjoyable collaborations. L.A.A.\ is partially
supported by the U.S.  National Science Foundation (NSF) under Grant
PHY-0757598 and CAREER Award PHY-1053663.  Any opinions, findings, and
conclusions or recommendations expressed in this material are those of
the author and do not necessarily reflect the views of the NSF.

\appendix

\section{T\lowercase{e}V-scale strings and large extra dimensions}
\label{tev}



\begin{figure}[tbp]
\postscript{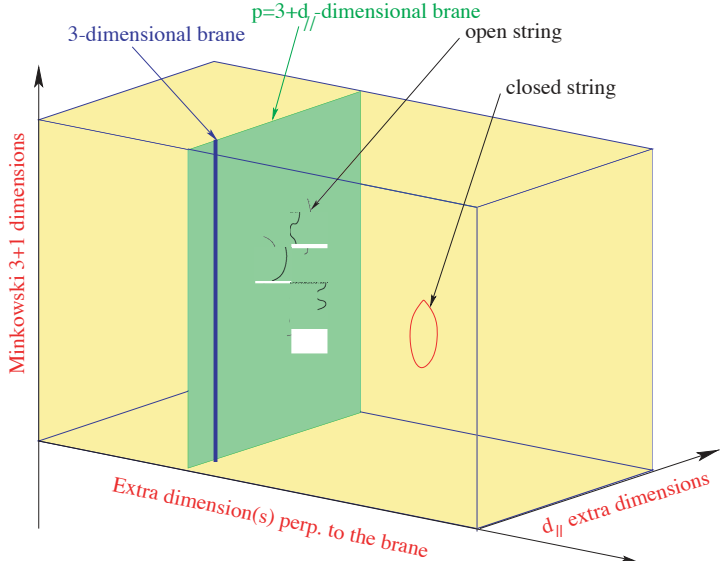}{0.6}
\caption[D-brane set-up]{D-brane
  set-up with $d_\parallel$ parallel and $d_\perp$ transverse internal
  directions. From Ref.~\cite{Antoniadis:2005aq}.}
\label{typeimodel}
\end{figure}

For an illustration, consider  type II string theory compactified on a six-dimensional torus $T^6$, which includes a D$p$-brane wrapped around $p-3$ dimensions of $T^6$ with the remaining dimensions along our familiar (uncompactified) three spatial dimensions. We denote the radii of the {\it internal} longitudinal directions (of the D$p$-brane) by $R_i^\parallel$, $i = 1, \dots p-3$ and the radii of the transverse directions by $R^\perp_j$, $j = 1, \dots 9-p$, see Fig.~\ref{typeimodel}.

The 4-dimensional Planck mass $M_{\rm Pl} \sim 10^{19}~{\rm GeV}$, which is related to the string mass scale $M_s$ by
\begin{equation}
M_{\rm Pl}^2=\frac{8}{g_s^2} \ M_s^8\
\frac{V_6}{(2\pi)^{6}}\ ,
\label{extradim}
\end{equation}
determines the strength of the gravitational interactions~\cite{Lust:2008qc}. Here,
\begin{equation}
V_6=(2\pi)^6\ \prod_{i=1}^{p-3} R^\parallel_i\ \prod_{j=1}^{9-p}
R^\perp_j
\end{equation}
is the volume of $T^6$ and $g_s$ is the string coupling. It follows
that the string scale can be chosen hierarchically smaller than the
Planck mass at the expense of introducing $n = 9-p$ large transverse
dimensions felt only by gravity, while keeping the string coupling
small. E.g. for a string mass scale $M_s \approx {\cal O} (1~{\rm TeV})$ the volume of the internal space needs to be as large as $V_6 M_s^6 \approx {\cal O} (10^{32}).$

On the other hand, the strength of coupling of the gauge theory
living on the D-brane world volume is not enhanced as long as
$R_i^\parallel \sim M_s^{-1}$ remain small, 
\begin{equation}
\frac{1}{g^2} = \frac{1}{2\pi \, g_s}\ M_s{}^{p-3}\ \
\prod_{i=1}^{p-3}R_i^\parallel\ .
\end{equation}
The weakness of the effective 4 dimensional gravity compared to gauge
interactions (ratio of $\langle H \rangle/M_{\rm Pl}$) is then attributed to the largeness of the transverse
space radii $R_i^\perp \sim 10^{32} l_s$ compared to the string length $l_s = M_s^{-1}$. 

A distinct property of these D-brane models is that gravity becomes
effectively $(4+n)$-dimensional with a strength comparable to those of
gauge interactions at the string scale~\cite{ArkaniHamed:1998rs}. Equation~(\ref{extradim}) can be
understood as a consequence of the $(4+n)$-dimensional Gauss law for
gravity, with
\begin{equation}
M_*^{2+n}  \sim \frac{1}{ g_s^2}  \, M_s^{2+n}
\end{equation}
the effective scale of gravity in $4+n$ dimensions. Taking $M_s \sim 1$~TeV, one finds a size for the extra dimensions $R^\perp \approx 10^{30/n-19}~{\rm m}.$ This relation immediately suggests that $n=1$ is ruled out, because $R^\perp \sim 10^{11}~{\rm m}$ and the gravitational interaction would thus be modified at the scale of our solar system. However, already for $n=2$ one obtains $R^\perp \sim 1~{\rm mm}$. This is just the scale where our present day experimental knowledge about gravity ends~\cite{Hoyle:2000cv}.

It is important to note that the mass scale $M_{\rm BH} \sim M_s/g_s^2$, which corresponds to the onset of black hole production, cannot be probed at the LHC ~\cite{Meade:2007sz}. To be specific, we choose $g_s = 0.1$ and then we obtain $M_{\rm BH} \sim 100 \, M_s$. It is also noteworthy that TeV-scale string D-brane compactifications naturally and unavoidably give rise to `the increidble bulk' which characterizes the recently proposed dynamical dark matter framework~\cite{Dienes:2011ja}.

\section{Pole residues of the Veneziano form factor}
\label{venpole}

Consider the product of Gamma functions
\begin{equation}
\Gamma(n) \, \, \Gamma(1-n) = \frac{\pi}{\sin (n\pi)} \ \  .
\end{equation}
In the limit  $1-n = \epsilon \ll 1$, 
$\sin (n \pi) = \sin (\pi - \pi \epsilon) = 
\sin (\pi) - \pi \epsilon \cos (\pi) 
= \pi \epsilon$, and so
\begin{equation}
\Gamma ( 1 - \epsilon) \,  \, \Gamma (\epsilon) = \frac{\pi}{\pi \epsilon} = \frac{1}{\epsilon} \, ,
\end{equation}
which in turn leads to
\begin{equation}
\lim_{n \to 1} \Gamma (1 - n) = \frac{1}{1-n} \, .
\end{equation} 
Therefore, in the limit of $s/M_s^2 \to 1$,
\begin{equation}
\mu (s, t, u) \to \frac{\Gamma (1 -u/M_s^2)}{\Gamma (1 + t/M_s^2)} \frac{1}{(1-s/M_s^2)} = 
\frac{\Gamma (2+t/M_s^2)}{\Gamma (1 + t/M_s^2)} \, \frac{1}{1-s/M_s^2} = \frac{1+t/M_s^2}{1+s/M_s^2} 
\label{label1}
\end{equation}
and
\begin{equation}
\mu (s,u,t) \to  \frac{\Gamma (1 -t/M_s^2)}{\Gamma(1 + u/M_s^2)} 
\frac{1}{(1-s/M_s^2)} =
\frac{\Gamma (2+u/M_s^2)}{\Gamma (1 + u/M_s^2)} \,\frac{1}{1-s/M_s^2}
= \frac{1+u/M_s^2}{1 +s/M_s^2} \, ;
\label{label2}
\end{equation}
recall we are working on the physical region (where $t$ and $u$ are
negatives) and so the second term in $\mu (s,u,t)$ or $\mu(s,t,u)$
does not develop Regge poles. We can now expand the string squared
amplitude,
\begin{eqnarray}
|{\cal M} (gg \to \gamma g)|^2 & \propto & 
\left| \frac{s}{u} \, \mu(s,t,u) + \frac{s}{t} \, \mu(s,u,t) \right|^2 
+ \left| \frac{t}{u} \, \mu(s,t,u) + \frac{t}{s} \, \mu(t,u,s) \right|^2 
\nonumber \\
& + & \left|\frac{u}{s} \, \mu(u,t,s) + \frac{u}{t} \, \mu(u,s,t) 
\right|^2 \,,
\end{eqnarray}
near the pole yielding
\begin{eqnarray}
|{\cal M} (gg \to \gamma g)|^2 & \propto & \left| \frac{M_s^2}{u} \, \, \frac{\Gamma(1-u/M_s^2)}{\Gamma(1+t/M_s^2)} + \frac{M_s^2}{t}  \, \, 
\frac{\Gamma(1-t/M_s^2)}{\Gamma(1+u/M_s^2)} \right|^2 \,
\frac{1}{(s - M_s^2)^2} \nonumber \\
& + & \left| -\frac{t}{u} \, \, \frac{\Gamma(1-u/M_s^2)}{\Gamma(1+t/M_s^2)} + 
\frac{t}{M_s^2} \, \, \left[ \frac{\Gamma(1-t/M_s^2)}{\Gamma(1+u/M_s^2)} - \frac{\Gamma(1-u/M_s^2)}{\Gamma(1+t/M_s^2)}\right] \right|^2 \, \, \frac{1}{(s - M_s^2)^2} \nonumber \\
   & + &  \left| \frac{u}{M_s^2} \, \, \left[ 
\frac{\Gamma(1-u/M_s^2)}{\Gamma(1+t/M_s^2)} -\frac{\Gamma(1-t/M_s^2)}{\Gamma(1+u/M_s^2)}\right] - \frac{u}{t}  \, \,  \frac{\Gamma(1-t/M_s^2)}{\Gamma(1+u/M_s^2)} \right|^2 \, \,
\frac{1}{(s - M_s^2)^2} \, . \nonumber 
\end{eqnarray} 
Equivalently,
\begin{eqnarray}
|{\cal M} (gg \to \gamma g)|^2 & \propto & \left\{ \left| \frac{M_s^2}{u} \ A + \frac{M_s^2}{t} \ B \right|^2 +\left|-\frac{t}{u}  \ A + \frac{t}{M_s^2} \ (B-A) \right|^2 \right. \nonumber \\ 
 & + & \left. \left|\frac{u}{M_s^2} \ (A-B) - \frac{u}{t} \ B \right|^2 \right\} 
\ \ \frac{1}{(s - M_s^2)^2} \, \,,
\label{ponyU}
\end{eqnarray}
where
\begin{equation}
A = \frac{\Gamma(1-u/M_s^2)}{\Gamma (1+t/M_s^2)} = 1 +t/M_s^2 = -u/M_s^2
\end{equation}
and
\begin{equation}
B = \frac{\Gamma(1-t/M_s^2)}{\Gamma(1+u/M_s^2)} = 1 + u/M_s^2 = -t/M_s^2 
\end{equation}
are obtained from Eqs.~(\ref{label1}) and (\ref{label2}).
Then, Eq.~(\ref{ponyU}) becomes
\begin{eqnarray}
|{\cal M} (gg \to \gamma g)|^2 & \propto & \left\{4 M_s^4 +  \left| t + \frac{t}{M_s^2} (-t + u)\right|^2 + \left| u + \frac{u}{M_s^2} (-u + t)\right|^2 \right\} \frac{1}{(s-M_s^2)^2} \nonumber \\
 & \propto &  \left[4 M_s^4 + \frac{4t^4 + 4u^4}{M_s^4}\right]  \,\, \frac{1}{(s-M_s^2)^2} \, \,,
\end{eqnarray}
where we have used the Mandelstam relation: $u = -M_s^2 -t$.


\begin{thebibliography}{99}



\bibitem{Nakamura:2010zzi}
  K.~Nakamura {\it et al.}  [Particle Data Group],
  J.\ Phys.\ G {\bf 37}, 075021 (2010).

\bibitem{Halzen:1984mc} For a thorough introduction to the SM, see for example,
  F.~Halzen and A.~D.~Martin,
``Quarks and Leptons: An Introductory Course in Modern Particle Physics,''  
(John Wiley \& Sons, New York, 1984)



\bibitem{GonzalezGarcia:2007ib}
  M.~C.~Gonzalez-Garcia and M.~Maltoni,
  Phys.\ Rept.\  {\bf 460}, 1 (2008)
  [arXiv:0704.1800 [hep-ph]].



\bibitem{Wess:1992cp}
  J.~Wess and J.~Bagger,
 ``Supersymmetry and Supergravity,'' 
2nd edition (Princeton University Press, Princeton, NJ, 1992).



\bibitem{Anchordoqui:2011eg}
  L.~A.~Anchordoqui, I.~Antoniadis, H.~Goldberg, X.~Huang, D.~Lust and T.~R.~Taylor,
  arXiv:1107.4309 [hep-ph].


\bibitem{Ibanez:2001nd}
  L.~E.~Ibanez, F.~Marchesano and R.~Rabadan,
  JHEP {\bf 0111}, 002 (2001)
  [arXiv:hep-th/0105155];
  D.~Cremades, L.~E.~Ibanez and F.~Marchesano,
  JHEP {\bf 0307}, 038 (2003)
  [arXiv:hep-th/0302105];
  F.~G.~Marchesano Buznego,
  arXiv:hep-th/0307252.


\bibitem{Peccei:1977hh}
  R.~D.~Peccei and H.~R.~Quinn,
  Phys.\ Rev.\ Lett.\  {\bf 38}, 1440 (1977);
  S.~Weinberg,
  Phys.\ Rev.\ Lett.\  {\bf 40}, 223 (1978);
  F.~Wilczek,
  Phys.\ Rev.\ Lett.\  {\bf 40}, 279 (1978).


\bibitem{Antoniadis:2000ena}
  I.~Antoniadis, E.~Kiritsis and T.~N.~Tomaras,
  Phys.\ Lett.\  B {\bf 486}, 186 (2000)
  [arXiv:hep-ph/0004214].

\bibitem{Anastasopoulos:2006da}
  P.~Anastasopoulos, T.~P.~T.~Dijkstra, E.~Kiritsis and A.~N.~Schellekens,
  Nucl.\ Phys.\  B {\bf 759}, 83 (2006)
  [arXiv:hep-th/0605226].





\bibitem{delAguila:1988jz}
  F.~del Aguila, G.~D.~Coughlan and M.~Quiros,
  Nucl.\ Phys.\  B {\bf 307}, 633 (1988)
  [Erratum-ibid.\  B {\bf 312}, 751 (1989)].


\bibitem{Haim} H.~Goldberg, unpublished.

\bibitem{Ghilencea:2002da}
  D.~M.~Ghilencea, L.~E.~Ibanez, N.~Irges and F.~Quevedo,
  JHEP {\bf 0208} (2002) 016 
  [arXiv:hep-ph/0205083].





\bibitem{Aaltonen:2011mk}
  T.~Aaltonen {\it et al.}  [CDF Collaboration],
  Phys.\ Rev.\ Lett.\  {\bf 106}, 171801 (2011)
  [arXiv:1104.0699 [hep-ex]];
  V.~Cavaliere,
  FERMILAB-THESIS-2010-51.



\bibitem{Punzi} G. Punzi, talk given at the {\em 23th Recontres de
    Blois,} Blois France, May, 2011. For the latest update of CDF
  analysis see A. Annovi, P. Catastini, V. Cavaliere, and L. Ristori,
  {\tt http://www-cdf.fnal.gov/physics/ewk/2011/wjj/7$_-$3.html}.


\bibitem{Abazov:2011af}
  V.~M.~Abazov {\it et al.}  [D\O\ Collaboration],
Phys.\ Rev.\ Lett.\  {\bf 107}, 011804 (2011)
  [arXiv:1106.1921 [hep-ex]].


\bibitem{Eichten:2011sh} 
  E.~J.~Eichten, K.~Lane and A.~Martin,
  Phys.\ Rev.\ Lett.\  {\bf 106}, 251803 (2011)
  [arXiv:1104.0976 [hep-ph]];
  C.~Kilic and S.~Thomas,
  arXiv:1104.1002 [hep-ph];
  A.~E.~Nelson, T.~Okui and T.~S.~Roy,
  arXiv:1104.2030 [hep-ph];
  B.~A.~Dobrescu and G.~Z.~Krnjaic,
  arXiv:1104.2893 [hep-ph];
 S.~Jung, A.~Pierce and J.~D.~Wells,
  arXiv:1104.3139 [hep-ph];
  M.~Buckley, P.~Fileviez Perez, D.~Hooper and E.~Neil,
  Phys.\ Lett.\  B {\bf 702}, 256 (2011)
  [arXiv:1104.3145 [hep-ph]];
 T.~Plehn and M.~Takeuchi,
  J.\ Phys.\ G {\bf 38}, 095006 (2011)
  [arXiv:1104.4087 [hep-ph]];
 Q.~H.~Cao, M.~Carena, S.~Gori, A.~Menon, P.~Schwaller, C.~E.~M.~Wagner and L.~T.~M.~Wang,
  JHEP {\bf 1108}, 002 (2011)
  [arXiv:1104.4776 [hep-ph]];
  J.~Fan, D.~Krohn, P.~Langacker and I.~Yavin,
  arXiv:1106.1682 [hep-ph];
  J.~L.~Evans, B.~Feldstein, W.~Klemm, H.~Murayama and T.~T.~Yanagida,
  arXiv:1106.1734 [hep-ph];
  R.~Harnik, G.~D.~Kribs and A.~Martin,
  arXiv:1106.2569 [hep-ph];
  J.~F.~Gunion,
  arXiv:1106.3308 [hep-ph].





\bibitem{Buckley:2011vc}
  M.~R.~Buckley, D.~Hooper, J.~Kopp and E.~Neil,
  Phys.\ Rev.\  D {\bf 83}, 115013 (2011)
  [arXiv:1103.6035 [hep-ph]];
  F.~Yu,
  Phys.\ Rev.\  D {\bf 83}, 094028 (2011)
  [arXiv:1104.0243 [hep-ph]];
 K.~Cheung and J.~Song,
  Phys.\ Rev.\ Lett.\  {\bf 106}, 211803 (2011)
  [arXiv:1104.1375 [hep-ph]];
  L.~A.~Anchordoqui, H.~Goldberg, X.~Huang, D.~Lust and T.~R.~Taylor,
  Phys. Lett. B {\bf 701}, 224 (2011)
  [arXiv:1104.2302 [hep-ph]];
  P.~Ko, Y.~Omura and C.~Yu,
  arXiv:1104.4066 [hep-ph];
  P.~J.~Fox, J.~Liu, D.~Tucker-Smith and N.~Weiner,
  arXiv:1104.4127 [hep-ph];
 Z.~Liu, P.~Nath and G.~Peim,
  Phys.\ Lett.\  B {\bf 701}, 601 (2011)
  [arXiv:1105.4371 [hep-ph]];
  M.~R.~Buckley, D.~Hooper and J.~L.~Rosner,
  arXiv:1106.3583 [hep-ph];
  A.~E.~Faraggi and V.~M.~Mehta,
  arXiv:1106.5422 [hep-ph];
  K.~Cheung and J.~Song,
  arXiv:1106.6141 [hep-ph].
See also Ref.~\cite{Anchordoqui:2011eg}.





 \bibitem{Acosta:2005ij}
  D.~E.~Acosta {\it et al.}  [CDF Collaboration],
  Phys.\ Rev.\ Lett.\  {\bf 95}, 131801 (2005)
  [arXiv:hep-ex/0506034];
  T.~Aaltonen {\it et al.}  [CDF Collaboration],
  Phys.\ Rev.\ Lett.\  {\bf 99}, 171802 (2007)
  [arXiv:0707.2524];
  T.~Aaltonen {\it et al.}  [CDF Collaboration],
  Phys.\ Rev.\ Lett.\  {\bf 102}, 091805 (2009)
  [arXiv:0811.0053].



\bibitem{Barate:1999qx} See {\it e.g.}
  R.~Barate {\it et al.}  [ALEPH Collaboration],
  Eur.\ Phys.\ J.\  C {\bf 12}, 183 (2000)
  [arXiv:hep-ex/9904011];
  P.~Abreu {\it et al.}  [DELPHI Collaboration],
  Phys.\ Lett.\  B {\bf 485}, 45 (2000)
  [arXiv:hep-ex/0103025];
  J.~Abdallah {\it et al.}  [DELPHI Collaboration],
  Eur.\ Phys.\ J.\  C {\bf 45}, 589 (2006)
  [arXiv:hep-ex/0512012].




\bibitem{Umeda:1998nq}
  Y.~Umeda, G.~C.~Cho and K.~Hagiwara,
  Phys.\ Rev.\  D {\bf 58}, 115008 (1998)
  [arXiv:hep-ph/9805447].




\bibitem{Abe:1993kb}
  F.~Abe {\it et al.}  [CDF Collaboration],
  Phys.\ Rev.\  D {\bf 48}, 998 (1993).
  F.~Abe {\it et al.}  [CDF Collaboration],
  Phys.\ Rev.\ Lett.\  {\bf 74}, 3538 (1995)
  [arXiv:hep-ex/9501001];
  F.~Abe {\it et al.}  [CDF Collaboration],
  Phys.\ Rev.\  D {\bf 55}, 5263 (1997)
  [arXiv:hep-ex/9702004];
  B.~Abbott {\it et al.}  [ D\O\ Collaboration],
  Phys.\ Rev.\ Lett.\  {\bf 82}, 2457 (1999)
  [arXiv:hep-ex/9807014];
  T.~Aaltonen {\it et al.}  [CDF Collaboration],
  Phys.\ Rev.\  D {\bf 79}, 112002 (2009)
  [arXiv:0812.4036].







\bibitem{Alitti:1990kw}
  J.~Alitti {\it et al.}  [UA2 Collaboration],
  Z.\ Phys.\  C {\bf 49}, 17 (1991);
  J.~Alitti {\it et al.}  [UA2 Collaboration],
  Nucl.\ Phys.\  B {\bf 400}, 3 (1993).





\bibitem{Hewett:2011nb}
  J.~L.~Hewett and T.~G.~Rizzo,
  arXiv:1106.0294 [hep-ph].

\bibitem{Williams:2011qb}
  M.~Williams, C.~P.~Burgess, A.~Maharana and F.~Quevedo,
  arXiv:1103.4556 [hep-ph].


\bibitem{Eichten:2011xd}
  E.~Eichten, K.~Lane and A.~Martin,
  arXiv:1107.4075 [hep-ph];
  K.~Harigaya, R.~Sato and S.~Shirai,
  arXiv:1107.5265 [hep-ph].



\bibitem{Buckley:2011hi}
  M.~R.~Buckley, D.~Hooper, J.~Kopp, A.~Martin and E.~T.~Neil,
  arXiv:1107.5799 [hep-ph].





\bibitem{Pumplin:2002vw}
  J.~Pumplin, D.~R.~Stump, J.~Huston, H.~L.~Lai, P.~Nadolsky and W.~K.~Tung,
  JHEP {\bf 0207}, 012 (2002)
  [arXiv:hep-ph/0201195].




\bibitem{Barger:1996kr}
  V.~D.~Barger, K.~M.~Cheung and P.~Langacker,
  Phys.\ Lett.\  B {\bf 381}, 226 (1996)
  [arXiv:hep-ph/9604298].


\bibitem{Kunszt:1992tn}
  Z.~Kunszt and D.~E.~Soper,
  Phys.\ Rev.\  D {\bf 46}, 192 (1992).


\bibitem{UA2-comment} Similar arguments have been previously advocated by  Nelson, Okui, and Roy in Ref.~\cite{Eichten:2011sh} and by Liu, Nath, and Peim in Ref.~\cite{Buckley:2011vc}.


\bibitem{Collaboration:2011dc}
G. Aad {\em et al.}  [ATLAS Collaboration],
  arXiv:1108.1582 [hep-ex].




\bibitem{Barger} V. Barger and R. J. N. Phillips, ``Collider Physics,''  
 (Addison-Wesley, 1987). 


\bibitem{Chatrchyan:2011ns}
  S.~Chatrchyan {\it et al.}  [CMS Collaboration],
  arXiv:1107.4771 [hep-ex].






\bibitem{Polchinski:1995mt}
  J.~Polchinski,
  Phys.\ Rev.\ Lett.\  {\bf 75}, 4724 (1995)
  [arXiv:hep-th/9510017];
  J.~Polchinski,
  arXiv:hep-th/9611050.


\bibitem{Antoniadis:1998ig}
  I.~Antoniadis, N.~Arkani-Hamed, S.~Dimopoulos and G.R.~Dvali,
  Phys.\ Lett.\  B {\bf 436}, 257 (1998)
  [arXiv:hep-ph/9804398].



\bibitem{Blumenhagen:2001te}
R.~Blumenhagen, B.~K\"ors, D.~L\"ust, T.~Ott,
  Nucl.\ Phys.\  B {\bf 616}, 3 (2001)
  [hep-th/0107138];
  M.~Cvetic, G.~Shiu and A.~M.~Uranga,
  Phys.\ Rev.\ Lett.\  {\bf 87}, 201801 (2001)
  [arXiv:hep-th/0107143];
  M.~Cvetic, G.~Shiu and A.~M.~Uranga,
  Nucl.\ Phys.\  B {\bf 615}, 3 (2001)
  [arXiv:hep-th/0107166];
  I.~Antoniadis, E.~Kiritsis and T.~Tomaras,
  Fortsch.\ Phys.\  {\bf 49}, 573 (2001)
  [arXiv:hep-th/0111269].
See also Ref.~\cite{Ibanez:2001nd}.

\bibitem{Kiritsis:2002aj} 
  E.~Kiritsis and P.~Anastasopoulos,
  JHEP {\bf 0205}, 054 (2002)
  [arXiv:hep-ph/0201295]; 
  G.~Honecker, T.~Ott,
  Phys.\ Rev.\  D {\bf  70}, 126010 (2004)
  [hep-th/0404055];
  F.~Gmeiner, R.~Blumenhagen, G.~Honecker, D.~Lust and T.~Weigand,
  JHEP {\bf 0601}, 004 (2006)
  [arXiv:hep-th/0510170];
  F.~Gmeiner, G.~Honecker,
  JHEP {\bf 0807}, 052 (2008)
  [arXiv:0806.3039 [hep-th]].


\bibitem{Kiritsis:2003mc} For reviews see {\em e.g.}
  E.~Kiritsis,
  Phys.\ Rept.\  {\bf 421}, 105 (2005)
  [Erratum-ibid.\  {\bf 429}, 121 (2006)]
  [Fortsch.\ Phys.\  {\bf 52}, 200 (2004)]
  [arXiv:hep-th/0310001];
  R.~Blumenhagen, M.~Cvetic, P.~Langacker and G.~Shiu,
  Ann.\ Rev.\ Nucl.\ Part.\ Sci.\  {\bf 55}, 71 (2005)
  [arXiv:hep-th/0502005].
  R.~Blumenhagen, B.~K\"ors, D.~L\"ust, S.~Stieberger,
  Phys.\ Rept.\  {\bf 445}, 1 (2007)
  [hep-th/0610327]. 
  

\bibitem{Antoniadis:2004dt}
  I.~Antoniadis and S.~Dimopoulos,
  Nucl.\ Phys.\  B {\bf 715} (2005) 120
  [arXiv:hep-th/0411032].

\bibitem{Berenstein:2006pk}
  D.~Berenstein and S.~Pinansky,
  Phys.\ Rev.\  D {\bf 75}, 095009 (2007)
  [arXiv:hep-th/0610104];
  D.~Berenstein, R.~Martinez, F.~Ochoa and S.~Pinansky,
  Phys.\ Rev.\  D {\bf 79}, 095005 (2009)
  [arXiv:0807.1126].


\bibitem{Green:1984sg}
  M.~B.~Green and J.~H.~Schwarz,
  Phys.\ Lett.\  B {\bf 149}, 117 (1984);
  E.~Witten,
  Phys.\ Lett.\  B {\bf 149}, 351 (1984);
  M.~Dine, N.~Seiberg and E.~Witten,
  Nucl.\ Phys.\  B {\bf 289}, 589 (1987);
  J.~J.~Atick, L.~J.~Dixon and A.~Sen,
  Nucl.\ Phys.\  B {\bf 292}, 109 (1987);
  W.~Lerche, B.~E.~W.~Nilsson, A.~N.~Schellekens and N.~P.~Warner,
  Nucl.\ Phys.\  B {\bf 299}, 91 (1988).



\bibitem{Antoniadis:2002cs}
  I.~Antoniadis, E.~Kiritsis and J.~Rizos,
  Nucl.\ Phys.\  B {\bf 637}, 92 (2002) 
  [arXiv:hep-th/0204153];
  P.~Anastasopoulos,
  JHEP {\bf 0308}, 005 (2003)
  [arXiv:hep-th/0306042].



  
  
 
  

\bibitem{Antoniadis:2002qm}
  I.~Antoniadis, E.~Kiritsis, J.~Rizos and T.~N.~Tomaras,
  Nucl.\ Phys.\  B {\bf 660}, 81 (2003) 
  [arXiv:hep-th/0210263].

\bibitem{Conlon:2008wa}
  J.~P.~Conlon, A.~Maharana, F.~Quevedo,
  JHEP {\bf 0905 }, 109 (2009).
  [arXiv:0810.5660 [hep-th]].
  

\bibitem{Mirjam}
  M.~Cvetic, J.~Halverson and P.~Langacker,
  arXiv:1108.5187 [hep-ph]. See also 
  M.~Cvetic, J.~Halverson and R.~2.~Richter,
  JHEP {\bf 0912}, 063 (2009)
  [arXiv:0905.3379 [hep-th]].




\bibitem{Parke:1986gb}
  S.~J.~Parke and T.~R.~Taylor,
  Phys.\ Rev.\ Lett.\  {\bf 56}, 2459 (1986).

\bibitem{Stieberger:2006te}
 S.~Stieberger and T. R.~Taylor,
  Phys.\ Rev.\ Lett.\  {\bf 97}, 211601 (2006)
  [arXiv:hep-th/0607184].
  S.~Stieberger and T.~R.~Taylor,
  Phys.\ Rev.\  D {\bf 74}, 126007 (2006)
  [arXiv:hep-th/0609175].




\bibitem{Mangano:1990by}
  M.~L.~Mangano and S.~J.~Parke,
  Phys.\ Rept.\  {\bf 200}, 301 (1991)
  [arXiv:hep-th/0509223];
  L.~J.~Dixon,
  arXiv:hep-ph/9601359.




\bibitem{Veneziano:1968yb}
  G.~Veneziano,
  Nuovo Cim.\  A {\bf 57}, 190 (1968).




\bibitem{Anchordoqui:2007da}
  L.~A.~Anchordoqui, H.~Goldberg, S.~Nawata and T.~R.~Taylor,
  Phys.\ Rev.\ Lett.\  {\bf 100}, 171603 (2008)
  [arXiv:0712.0386 [hep-ph]].

\bibitem{vanRitbergen:1998pn}
  T.~van Ritbergen, A.~N.~Schellekens and J.~A.~M.~Vermaseren,
  Int.\ J.\ Mod.\ Phys.\  A {\bf 14}, 41 (1999)
  [arXiv:hep-ph/9802376].

\bibitem{Lust:2008qc}
  D.~L\"ust, S.~Stieberger and T.~R.~Taylor,
  Nucl.\ Phys.\  B {\bf 808}, 1 (2009)
  [arXiv:0807.3333 [hep-th]].

\bibitem{Anchordoqui:2008ac}
  L.~A.~Anchordoqui, H.~Goldberg, S.~Nawata and T.~R.~Taylor,
  Phys. Rev. D {\bf 78}, 016005 (2008)
  [arXiv:0804.2013 [hep-ph]].




\bibitem{Anchordoqui:2008di}
  L.~A.~Anchordoqui, H.~Goldberg, D.~L\"ust, S.~Nawata, S.~Stieberger and T.~R.~Taylor,
  Phys.\ Rev.\ Lett.\  {\bf 101}, 241803 (2008)
  [arXiv:0808.0497 [hep-ph]].


\bibitem{Anchordoqui:2008hi}
  L.~A.~Anchordoqui, H.~Goldberg and T.~R.~Taylor,
  Phys.\ Lett.\  B {\bf 668}, 373 (2008)
  [arXiv:0806.3420 [hep-ph]].


\bibitem{Dong:2010jt}
  Z.~Dong, T.~Han, M.~x.~Huang and G.~Shiu,
  JHEP {\bf 1009}, 048 (2010)
  [arXiv:1004.5441 [hep-ph]].

\bibitem{Cornet:2001gy}
  F.~Cornet, J.~I.~Illana and M.~Masip,
  Phys.\ Rev.\ Lett.\  {\bf 86}, 4235 (2001)
  [arXiv:hep-ph/0102065].




\bibitem{Eichten:1984eu}
  E.~Eichten, I.~Hinchliffe, K.~D.~Lane and C.~Quigg,
  Rev.\ Mod.\ Phys.\  {\bf 56}, 579 (1984)
  [Addendum-ibid.\  {\bf 58}, 1065 (1986)].



\bibitem{Khachatryan:2010jd}
  V.~Khachatryan {\it et al.}  [CMS Collaboration],
  Phys.\ Rev.\ Lett.\  {\bf 105}, 211801 (2010)
  [Phys.\ Rev.\  {\bf 106}, 029902 (2011)]
  [arXiv:1010.0203 [hep-ex]].


\bibitem{Bhatti:2008hz}
  A.~Bhatti {\it et al.},
  J.\ Phys.\ G {\bf 36}, 015004 (2009)
  [arXiv:0807.4961 [hep-ex]].


\bibitem{Randall:1999ee}
  L.~Randall and R.~Sundrum,
  Phys.\ Rev.\ Lett.\  {\bf 83}, 3370 (1999)
  [arXiv:hep-ph/9905221].


\bibitem{Hassanain:2009at}
  B.~Hassanain, J.~March-Russell and J.~G.~Rosa,
  JHEP {\bf 0907}, 077  (2009)
  [arXiv:0904.4108 [hep-ph]];
  M.~Perelstein and A.~Spray,
  JHEP {\bf 0910}, 096 (2009)
  [arXiv:0907.3496 [hep-ph]];
L.~A.~Anchordoqui, H.~Goldberg, X.~Huang and T.~R.~Taylor,
  Phys.\ Rev.\  D {\bf 82}, 106010 (2010)
  [arXiv:1006.3044 [hep-ph]];
  M.~Perelstein and A.~Spray,
  arXiv:1106.2171 [hep-ph].









\bibitem{Anchordoqui:2009ja}
  L.~A.~Anchordoqui, H.~Goldberg, D.~Lust, S.~Stieberger and T.~R.~Taylor,
  Mod.\ Phys.\ Lett.\  A {\bf 24}, 2481 (2009)
  [arXiv:0909.2216 [hep-ph]].


\bibitem{Ball:2007zza}
  G.~L.~Bayatian {\it et al.}  [CMS Collaboration],
  J.\ Phys.\ G {\bf 34} 995 (2007);
  W.~W.~Armstrong {\it et al.}  [ATLAS Collaboration],
  CERN/LHCC 94-43.



\bibitem{Gupta:2007cy}
  P.~Gupta, B.~C.~Choudhary, S.~Chatterji, S.~Bhattacharya and R.~K.~Shivpuri,
  arXiv:0705.2740 [hep-ex].



\bibitem{Cullen:2000ef}
  S.~Cullen, M.~Perelstein and M.~E.~Peskin,
  Phys.\ Rev.\  D {\bf 62}, 055012 (2000)
  [arXiv:hep-ph/0001166].
 




\bibitem{Anchordoqui:2009mm}
  L.~A.~Anchordoqui, H.~Goldberg, D.~Lust, S.~Nawata, S.~Stieberger and T.~R.~Taylor,
  Nucl.\ Phys.\  B {\bf 821}, 181 (2009)
  [arXiv:0904.3547 [hep-ph]].



\bibitem{Abbott:1998wh}
  B.~Abbott {\it et al.}  [ D\O\ Collaboration],
  Phys.\ Rev.\ Lett.\  {\bf 82}, 2457 (1999)
  [arXiv:hep-ex/9807014].


\bibitem{Meade:2007sz}  
  P.~Meade and L.~Randall,
  JHEP {\bf 0805}, 003 (2008)
  [arXiv:0708.3017 [hep-ph]].






\bibitem{Esen}
  S. Esen and R. Harris,
  CMS Note 2006/071.



\bibitem{Anchordoqui:2010zs}
  L.~A.~Anchordoqui, W.~Z.~Feng, H.~Goldberg, X.~Huang and T.~R.~Taylor,
  Phys.\ Rev.\  D {\bf 83}, 106006 (2011)
  [arXiv:1012.3466 [hep-ph]].











\bibitem{Antoniadis:2005aq}
  I.~Antoniadis,
  Lect.\ Notes Phys.\  {\bf 720}, 293 (2007)
  [arXiv:hep-ph/0512182].


\bibitem{ArkaniHamed:1998rs}
  N.~Arkani-Hamed, S.~Dimopoulos and G.~R.~Dvali,
  Phys.\ Lett.\  B {\bf 429}, 263 (1998)
  [arXiv:hep-ph/9803315].


\bibitem{Hoyle:2000cv}
  C.~D.~Hoyle, U.~Schmidt, B.~R.~Heckel, E.~G.~Adelberger, J.~H.~Gundlach, D.~J.~Kapner and H.~E.~Swanson,
  Phys.\ Rev.\ Lett.\  {\bf 86}, 1418 (2001)
  [arXiv:hep-ph/0011014];
  C.~D.~Hoyle, D.~J.~Kapner, B.~R.~Heckel, E.~G.~Adelberger, J.~H.~Gundlach, U.~Schmidt and H.~E.~Swanson,
  Phys.\ Rev.\  D {\bf 70}, 042004 (2004)
  [arXiv:hep-ph/0405262];
  D.~J.~Kapner, T.~S.~Cook, E.~G.~Adelberger, J.~H.~Gundlach, B.~R.~Heckel, C.~D.~Hoyle and H.~E.~Swanson,
  Phys.\ Rev.\ Lett.\  {\bf 98}, 021101 (2007)
  [arXiv:hep-ph/0611184].


\bibitem{Dienes:2011ja}
  K.~R.~Dienes and B.~Thomas,
  arXiv:1106.4546 [hep-ph];
  K.~R.~Dienes and B.~Thomas,
  arXiv:1107.0721 [hep-ph].






\end{thebibliography}
\end{document}